\documentclass[aps,prd,onecolumn,groupedaddress,showpacs,nofootinbib,amssymb,balancelastpage]{revtex4}
\usepackage[dvipdfmx]{graphicx}
\usepackage{graphicx,bm,color}

\usepackage{amsmath}
\usepackage{amssymb}
\usepackage{amsfonts}
\usepackage{cases}
\usepackage{cancel}
\usepackage{hyperref}
\usepackage[normalem]{ulem}
\usepackage{subfigure}
\usepackage{here}
\usepackage{color}
\usepackage{comment}
\usepackage{mathrsfs}

\usepackage{tikz}
\usetikzlibrary{matrix}

\allowdisplaybreaks[3]

\newcommand{\nLk}{n_{L, \vec{k}}^s}
\newcommand{\nRk}{n_{R, \vec{k}}^s}
\newcommand{\DLk}{D_{L, \vec{k}}^s}
\newcommand{\DRk}{D_{R, \vec{k}}^s}
\newcommand{\nLmk}{n_{L, -\vec{k}}^s}
\newcommand{\nRmk}{n_{R, -\vec{k}}^s}
\newcommand{\DLmk}{D_{L, -\vec{k}}^s}
\newcommand{\DRmk}{D_{R, -\vec{k}}^s}

\newcommand{\nLkd}{n_{L, \vec{k}}^{s \dagger}}
\newcommand{\nRkd}{n_{R, \vec{k}}^{s \dagger}}
\newcommand{\DLkd}{D_{L, \vec{k}}^{s \dagger}}
\newcommand{\DRkd}{D_{R, \vec{k}}^{s \dagger}}
\newcommand{\nLmkd}{n_{L, -\vec{k}}^{s \dagger}}
\newcommand{\nRmkd}{n_{R, -\vec{k}}^{s \dagger}}
\newcommand{\DLmkd}{D_{L, -\vec{k}}^{s \dagger}}
\newcommand{\DRmkd}{D_{R, -\vec{k}}^{s \dagger}}
\newcommand{\eiqk}{e^{i \theta_{\vec{k}}}}
\newcommand{\eiqmk}{e^{-i \theta_{\vec{k}}}}
\newcommand{\abk}{\left| \vec{k} \right|}
\newcommand{\tilmN}{\tilde{m}_N}

\begin{document}

\title{Baryogenesis via QCD preheating with nonadiabatic baryon chemical potential}   
\author{Jimin Wang}\thanks{{\tt jmwang22@mails.jlu.edu.cn}}
\affiliation{Center for Theoretical Physics and College of Physics, Jilin University, Changchun, 130012,
China.}

\author{Xin-Ru Wang}\thanks{{\tt wxr21@mails.jlu.edu.cn}}
\affiliation{Center for Theoretical Physics and College of Physics, Jilin University, Changchun, 130012,
China.}

\author{Shinya Matsuzaki}\thanks{{\tt synya@jlu.edu.cn}}
\affiliation{Center for Theoretical Physics and College of Physics, Jilin University, Changchun, 130012,
China.}

\begin{abstract}
The chiral phase transition in QCD can be supercooled in the thermal history of the universe to be instantaneously out-of equilibrium,  
if QCD is coupled to a dark QCD sector exhibiting the dark chiral phase transition of the first order.  In that case the QCD sigma meson field (as the chiral order parameter, or the light quark condensate) starts to roll in a nonadiabatic way down to the true QCD vacuum. 
Meanwhile a dynamic baryonic chemical potential can be generated solely within QCD, 
which is governed by the dynamic motion of the QCD sigma meson field, analogously to the spontaneous baryogenesis or the leptogenesis via the Higgs or axionlike relaxation scenario. When QCD is further allowed to communicate with a dark fermion with mass of order of 1 GeV and the baryon number violating coupling to neutron, the nonadiabatic QCD sigma motion along with the nonadiabatic baryon chemical potential can trigger the preheating and produce 
the baryon number asymmetry. 
We discuss this scenario in details to find that the QCD-induced dynamic baryon chemical potential plays a significant role for the QCD preheating and the baryogenesis, which yields the desired amount of the asymmetry today consistently with current astrophysical, cosmological, and terrestrial experimental constraints. 
Cosmological and phenomenological consequences characteristic to 
the present scenario are also addressed. 
\end{abstract}

\maketitle

\section{Introduction}

The cosmic microwave background (CMB) and the big bang nucleosynthesis (BBN) imply existence of the baryon number asymmetry (BNA) today, which is captured in terms of the ratio of the baryon number density to the entropy density with the amount $(n_B- \bar{n}_B)/s \sim 6 \times 10^{-10}$~\cite{Planck:2015fie,Cyburt:2015mya,Planck:2018vyg}, indicating 
the dominance of matters against anti-matters. 
The Standard Model (SM) of particle physics cannot sufficiently produce 
the present-day BNA, so this BNA issue is one of well motivated slogans for theoretical particle physicists to call for beyond the SM.

The mechanism to generate this asymmetry, 
called the baryogenesis, needs to satisfy the Sakharov's criteria~\cite{Sakharov:1967dj}: C and CP violation, baryon number violation, and out-of-equilibrium. 
Among the Sakharov's criteria, in particular, realization of non-equilibrium 
can involve a cosmological aspect, i.e., a dynamic and nonadiabatic time evolution of the vacuum. 
Such a dynamic aspect of the vacuum 
has been discussed aiming at the post inflationary universe, so-called preheating~\cite{Dolgov:1989us,Traschen:1990sw,Kofman:1994rk,Shtanov:1994ce,Kofman:1997yn} 
(for reviews, see e.g., \cite{Kofman:1997yn,Amin:2014eta,Lozanov:2019jxc}). 
When the inflaton couples to the SM particles, the nonadiabatic motion or oscillation of the inflaton generates the nonadiabatic quantum states for the SM particles, 
that is what is called the nonperturbative particle production.

Still, the dynamic vacuum can be transported from 
the baryon or lepton currents via higher dimensional operators induced from Beyond the SM, 
which yields the baryonic or leptonic chemical potential term, like 
$\sim \dot{\phi} n_{B/L}$ or $\dot{\phi^2} n_{B/L}$, where $n_{B/L}$ denotes 
the baryon (lepton) number density and $\phi$ a scalar field which determines the vacuum. The time-reversal invariance is also locally violated due to the time-dependent vacuum, so is the C and CP invariance. 
The original work on the spontaneous thermal baryogenesis~\cite{Cohen:1987vi} was based on 
this possibility, and recently it has been coupled to the preheating mechanism driven 
by the oscillating SM Higgs or the axionlike particles, to be encoded with scenarios such as 
the leptogenesis via the axion(like) scalar oscillation~\cite{Kusenko:2014uta,Ibe:2015nfa,Daido:2015gqa,Takahashi:2015waa,Adshead:2015jza,Takahashi:2015ula,Kusenko:2016vcq,Maleknejad:2016dci,DeSimone:2016ofp,DeSimone:2016juo,Son:2018avk,Dasgupta:2018eha,Bae:2018mlv,Domcke:2019qmm,Wu:2020pej,Berbig:2023uzs}, and the Higgs relaxation or  inflation~\cite{Kusenko:2014lra,Yang:2015ida,Pearce:2015nga,Gertov:2016uzs,Kawasaki:2017ycl,Wu:2019ohx,Lee:2020yaj,Cado:2021bia}. 
All those scenarios proposed so far have assumed the preheating at higher scales with the prediction of new particles heavier than the SM particles.

In this paper, we explore a lower-scale preheating with a dynamic and nonadiabatic 
baryon chemical potential: it is triggered at the QCD scale of 1 GeV. 
This is sort of what is called the QCD preheating, which has been recently proposed in the literature~\cite{Wang:2022ygk}. 
The presently addressed scenario is thus alternative to 
the original one, in a sense that it is the case when the QCD-induced baryon chemical potential is assumed to be sizable.

The QCD preheating with the dynamical baryon chemical potential is supposed 
to be more likely to take place, if the particle production in the universe would be nonadiabatic. The reason consists of two ingredients: i) 
as has been emphasized in~\cite{Wang:2022ygk}, 
a scalar condensate is present naturally in QCD, that is 
the light quark condensate $\sigma \sim \langle \bar{q}q \rangle $, 
and it can couple to the nucleon state as well as meson states 
in a systematic way respecting the chiral (or isospin) symmetry and its breaking 
by the current quark masses; ii) QCD potentially generates the coupling between 
the dynamic $\sigma$ field to the baryon current in the chiral (or isospin) symmetric way. This is indeed the setup that is assumed in higher-scale preheating scenarios with beyond the SM contents or higher dimensional operators induced from some beyond the SM. It is all built-in in the QCD preheating case.

An instantaneous out-of-equilibrium can be achieved 
if QCD is coupled to a dark QCD (dQCD) sector exhibiting the dark chiral phase transition of the first order, so that the chiral phase transition in QCD can be supercooled.  
 In that case the QCD sigma meson field (as the chiral order parameter, or the light quark condensate) starts to roll in a nonadiabatic way down to the true QCD vacuum. 
The induced dynamic baryonic chemical potential is governed by the dynamic motion of the QCD sigma meson field, so the preheating takes place along with the dynamic baryon chemical potential: out-of-equilibrium and CP violation are realized.

We assume QCD to be further allowed to communicate with a dark fermion with mass of order of 1 GeV and the baryon number violating coupling to neutron,  
then the QCD preheating produce the BNA. 
We discuss this scenario in details to find that the QCD-induced dynamic baryon chemical potential plays a significant role for the QCD preheating and the baryogenesis. 
We show that the desired amount of the asymmetry today is generated consistently with current astrophysical, cosmological, and terrestrial experimental constraints.

We also discuss and give comments on the cosmological and phenomenological consequences characteristic to 
the present scenario, such as the impacts on the QCD hadron physics, the cosmological and collider physics probes for the dQCD sector, and productions of the gravitational waves arising from the supercooled dQCD phase transition, 
including comparison between the present and previous QCD preheating scenarios.

This paper is structured as follows. 
In Sec.~\ref{sec2}, we start with introducing a low-energy effective model 
of QCD including the QCD sigma meson, pions, nucleons, and  
the induced dynamic chemical potential operator, which will be preliminary setups for 
the later discussions. 
In Sec.~\ref{sec3} we first demonstrate that thermal QCD baryogenesis, with assuming some 
Beyond the SM violating the baryon number symmetry, yields too small BNA. 
Then, in Sec.~\ref{sec4}, we turn to the preheating scenario driven by 
the supercooling in the dQCD sector coupled to QCD. Cosmological and phenomenological constraints on presence of the dQCD sector are addressed there.   
The detailed setups relevant to the preheating analysis are presented 
in Sec.~\ref{sec5}.  
We show the numerical results on the QCD preheating with the dynamic baryon chemical 
potential, in Sec.~\ref{sec6}, and see that the desired size of the BNA can be produced due to the QCD-induced nonadiabatic baryon chemical potential.  
Section~\ref{sec:sumamry} provides the summary of the present paper and 
several discussions related to the future prospected probes of the present preheating 
scenario. Supplemental formulae related to the numerical analysis on the preheating are listed in Appendices~\ref{BT} and \ref{two-point}.

\section{The LSM with induced dynamic baryon chemical potential}  
 \label{sec2} 

In this section we start with introducing a low-energy effective description  
of QCD including the QCD sigma meson, pions, nucleons, and  
the induced dynamic chemical potential operator, which setups the preliminaries in light of the discussions in the later sections. 
We employ the linear sigma model (LSM) with the lightest two flavors $q=(u,d)^T$ 
to monitor the low-energy chiral dynamics in QCD, based on the chiral $SU(2)_L \times SU(2)_R$ symmetry and its breaking structure. 
The building blocks are: 
two-by-two complex scalar matrix: 
$M \sim \bar{q}_R q_L$, which transforms 
under the chiral symmetry as $M \to U_L \cdot M \cdot U_R^\dag$ with  
$ U_{L,R} \in SU(2)_{L,R}$, 
where $M$ is parametrized by the isosinglet sigma ($\sigma$) mode 
and isotriplet pion ($\pi$) mode as 
$M = \sigma\cdot 1_{\bf 2 \times 2}/2 + i \pi^a \tau^a/2$ with the Pauli matrices $\tau^a$ ($a= 1,2,3$); nucleon (proton, neutron)-doublet field $N_{L,R} = (p, n)^T_{L,R}$, which belong to the fundamental representation of $SU(2)_{L,R}$ groups.

The LSM Lagrangian is thus given as 
\begin{align} 
& \mathcal{L}_{\rm LSM} 
= \operatorname{tr}\left[\partial_{\mu} M^{\dagger} \partial^{\mu} M\right] 
- V 
\notag\\ 
& \hspace{20pt} 
+\bar{N} i \gamma^\mu \partial_\mu N-\frac{2 m_{N}}{f_{\pi}}\left(\bar{N}_{L} M N_{R}+\bar{N}_{R} M^{\dagger} N_{L}\right) 
\,, \notag\\ 
& V  = 
m_{\pi}^{2} f_{\pi} \operatorname{tr}\left[{\rm Re}(M) \right] 
+ m^{2} \operatorname{tr}\left[M^{\dagger} M\right] 
+ \lambda\left(\operatorname{tr}\left[M^{\dagger} M\right]\right)^{2}
\,. \label{LSM-Lagrangian:eq}
\end{align}
All the terms are invariant under the chiral symmetry, 
C and P symmetries, except for the first term in $V$ coupled to  
the pion mass, which breaks the 
chiral symmetry in a way reflecting the current 
quark mass term in the underlying QCD. 
The explicit-chiral or -isospin breaking for the nucleon sector has been neglected for simplicity.

The light quark condensate $\langle \bar{q} q \rangle$ generated in QCD is aligned to be two-flavor diagonal (isospin symmetric) $SU(2)_V$ and P invariant,  so that $\langle - \bar{q}_i q_j \rangle \sim \Lambda_{\rm QCD}^3 \cdot \delta_{ij}$, which triggers 
the (approximate) spontaneous chiral symmetry breaking 
$SU(2)_L \times SU(2)_R \to SU(2)_V$ at the vacuum. 
Hence the dynamics of the vacuum expectation value of $\sigma $, 
$\langle \sigma \rangle$,  monitors the dynamic $\langle \bar{q} q \rangle$.

In Eq.(\ref{LSM-Lagrangian:eq}) and throughout the present study, 
we take the pion decay constant $f_\pi \simeq 92.4$ MeV, 
the (isospin symmetric) pion mass $m_\pi \simeq 140$ MeV,
and the (isospin symmetric nucleon mass) $m_N \simeq 940$ MeV. 
We fix the potential parameters $m^2$ and $\lambda$ in $V$ to the values determined 
at the true vacuum $\langle \sigma \rangle = f_\pi$ satisfying the stationary 
condition $\partial V/\partial \sigma |_{\sigma =f_\pi} =0$. 
Then we have $\lambda f_\pi^2 = (m_\sigma^2 - m_\pi^2)/2 $ 
and $m^2= \frac{3}{2} m_\pi^2  - \frac{1}{2}m_\sigma^2$, with the $\sigma$ mass squared defined as 
$m_\sigma^2 = \partial^2 V/\partial \sigma^2 |_{\sigma = f_\pi}$, which we take  ${\color{black}\simeq(500 \: {\rm MeV})^2}$ as the reference mass value of $f_0(500)$.

The vacuum expectation value of $\sigma$, $\langle \sigma \rangle$, and its dynamical evolution are identical to those for $\langle \bar{q}q \rangle$. 
In particular, the dynamic $\langle \sigma \rangle$ controls the mass of the nucleons through the Yukawa interaction in Eq.(\ref{LSM-Lagrangian:eq}): 
\begin{equation}
 \frac{2m_N}{f_\pi}\bar{N}_LMN_R = \frac{m_N}{f_\pi}\langle\sigma\rangle\cdot \bar{N}_LN_R+\cdots.
\label{mN:sigma}
\end{equation}
Therefore, the nucleon mass $\tilde{m}_N(t)=m_N\frac{\langle\sigma(t)\rangle}{f_\pi}$ varies in time, following the time evolution of $\langle\sigma\rangle$.

We further consider the following higher dimensional interaction, which can be induced from QCD and plays a role of the dynamic baryon chemical potential: 
%
%
\begin{align} 
 {\cal L}_{\mu_{\rm dyn}} =- \frac{c}{(4 \pi f_\pi)^2} \cdot \partial_\mu 
 {\rm tr}[M^\dag M] J^\mu_B
\,,  
\end{align}
where $J_B^\mu$ is the $U(1)$ baryon number current, 
$J_B^\mu = \bar{N} \gamma^\mu N$, and the coupling constant $c$ is of ${\cal O}(1)$ 
based on the naive dimensional analysis~\footnote{
Since the chemical potential operator is chiral and QCD gauge invariant, and constructed from the conserved current, it will not get renormalized, hence no nontrivial enhancement or suppression of the operator will be generated at the low-energy scale $\lesssim 4 \pi f_\pi \sim 1$ GeV.  
}. 
In the spatially homogeneous universe, ${\cal L}_{\mu_{\rm dyn}}$ 
leaves only an effective chemical potential term dependent of $M(t)$: 
\begin{align} 
  {\cal L}_{\mu_{\rm dyn}}  = - 
  \mu_{\rm dyn}(t) \cdot n_B(t)
\,, \qquad 
\mu_{\rm dyn}(t) \equiv \frac{c}{(4 \pi f_\pi)^2} \partial_0 
 {\rm tr}[M(t)^\dag M(t)] 
 = \frac{c}{32 \pi^2 f_\pi^2} \frac{d}{dt} \langle \sigma^2(t) \rangle 
\,, \label{mudyn}
\end{align} 
where $n_B(t) = N^\dag(t) N(t)$ is the baryon number density. 
When the dynamic $\mu_{\rm dyn}(t)$ gets 
activated, 
nonzero dynamic $\mu_{\rm dyn}(t)$ breaks the C and CP symmetries as well as the time-reversal $T$ symmetry. 
The form of the dynamic chemical potential operator $\sim \partial_0 {\rm tr} [\langle M^\dag M \rangle]$, not including the Nambu-Goldstone  boson mode (i.e. pions), is in contrast to the so-called spontaneous baryogenesis scenario~\cite{Cohen:1987vi}, or the leptogenesis via the axion(like) scalar oscillation~\cite{Kusenko:2014uta,Ibe:2015nfa,Daido:2015gqa,Takahashi:2015waa,Adshead:2015jza,Takahashi:2015ula,Kusenko:2016vcq,Maleknejad:2016dci,DeSimone:2016ofp,DeSimone:2016juo,Son:2018avk,Dasgupta:2018eha,Bae:2018mlv,Domcke:2019qmm,Wu:2020pej,Berbig:2023uzs}, which is rather analogous to a class of   
the leptogenesis via the Higgs relaxation or inflation~\cite{Kusenko:2014lra,Yang:2015ida,Pearce:2015nga,Gertov:2016uzs,Kawasaki:2017ycl,Wu:2019ohx,Lee:2020yaj,Cado:2021bia}.

\section{Thermal QCD-crossover baryogenesis}  
\label{sec3}

First of all, we shall consider the ordinary QCD thermal history, in as sense that 
$\langle \sigma \rangle \sim \langle \bar{q}q \rangle$ starts to {\it smoothly} roll from $\langle \sigma \rangle \sim 0$ 
(in the (nearly) chiral symmetric phase) down to the true vacuum at $\langle \sigma \rangle 
= f_\pi$ during the chiral crossover transition epoch, $0.7 T_{\rm pc} \lesssim T \lesssim T_{\rm pc}$ with the pseudo critical temperature $\sim 155$ MeV~\cite{Aoki:2009sc}. 
Once a sufficient amount of the baryon number violation is supplied from a dark sector beyond the SM, 
the thermal QCD baryogenesis would be processed via the dynamic $\langle \sigma(t) \rangle$ $(0.7 T_{\rm pc} \lesssim T \lesssim T_{\rm pc}$) until $\sigma(t)$ reaches the true vacuum $\langle \sigma \rangle = f_\pi$. 
Here we assume that 
the time variation of $ \mu_{\rm dyn}(t) $, hence the velocity $\dot{\sigma}(t)$ is slow enough that 
the baryon number violating process can still maintain the thermal equilibrium until $t=t_D$, and, when decoupled at $t=t_D$,  
yield the net asymmetry supplied safely via the thermal chemical potential $ \mu_{\rm dyn}(t_D) $.

The dynamic chemical potential $\mu_{\rm dyn}(t)$ thus  
 creates the initial net baryon number $n_B(0)$ arising via the 
 thermal distributions for the nonrelativistic nucleon $N$ and anti-nucleon $\bar{N}$, 
 which can be estimated, to the first nontrivial order in expansion with respect to 
$\mu_{\rm dyn}/T \ll 1$, as 
\begin{align}
    n_B(0) &\equiv n_N - n_{\bar{N}} 
    \notag\\ 
& = \frac{g_N}{2\pi^2} \int_{m_N}^\infty 
dE \, E (E^2- m_N^2)^{1/2} 
\left[ \frac{1}{1+ e^{\frac{E-\mu_{\rm dyn}}{T}}}
-\frac{1}{1+ e^{\frac{E+\mu_{\rm dyn}}{T}}} 
\right]
\notag\\ 
    & \simeq \frac{2 g_N}{(2 \pi)^{3/2}} 
 (m_N T)^{3/2} e^{- m_N/T} \sinh{\frac{\mu_{\rm dyn}(t)}{T}}
 \,, \label{nB0}
 \end{align} 
where $g_N$ denotes the degrees of freedom for the nucleon.  
 The initial thermal net number $n_B(0)$ in Eq.(\ref{nB0}) goes to zero, as $\mu_{\rm dyn} \to 0$. 
However, if the baryon number violating process decouples from 
the thermal equiliburium faster than $\mu_{\rm dyn}(t)$ reaches zero  (i.e. $\langle \sigma(t) \rangle$ reaches the true vacuum at $f_\pi$), the net number is frozen out at the decoupling temperature $T_D$, to be left today.

 In the case of the spontaneous baryogenesis, 
 a pseudo Nambu-Goldstone boson (coupled to the spontaneously broken baryon current) is 
assumed to be light so that the mass (i.e., inverse of the damping oscillation time scale) $\ll H$,  
--- the temperature of the Universe is assumed to be much higher than QCD scale as well ---  
hence the damping oscillation dynamics of the so-called {\it thermalion} significantly 
contributes to the thermal baryogenesis (critically assuming the thermalion to be
initially trapped at a slope of the potential)~\cite{Cohen:1987vi}.  
In contrast, 
the thermal QCD phase transition  
is crossover as observed in lattice simulations, 
so that $\langle \sigma(t) \rangle$ is simply expected to smoothly travel the vacuum transition from $\sim 0$ (at $T= T_{\rm pc}$) to $f_\pi$ (at $T \sim 0.7 T_{\rm pc}$) 
which evolves along with the Hubble expansion in an adiabatic manner. 
In that case there is no drastic trapping at (around) the origin of the potential 
in contrast to the literature~\cite{Cohen:1987vi}. 
The adiabatic motion of $\langle \sigma(t) \rangle$, 
which evolves along with the Hubble background, 
is thus relevant in the present scenario.

The number density per comoving volume 
$Y_B = n_B/s$ is adiabatically conserved until today: $n_B(0)/s(T_D) =  n_B(T_0)/s(T_0)$ 
with the thermal entropy densities $s(T) = \frac{2 \pi^2}{45} g_{*s}(T) T^3$. 
We refer to the crossover data on $\langle \sigma(T) \rangle $ from the lattice QCD 
for $0.7 T_{\rm pc} \lesssim T \lesssim T_{\rm pc}$~\cite{Aoki:2009sc} and 
choosing $T_D$ in the crossover regime:  
$T_D = {\cal O}( \langle \sigma \rangle)  
\lesssim {\cal O}(f_\pi)$ for $0.7 T_{\rm pc} \lesssim T \lesssim T_{\rm pc}$, 
$\langle \dot{\sigma} \rangle = {\cal O}(f_\pi^2)$, hence $\mu_{\rm dyn} = {\cal O}(c \cdot f_\pi/(4\pi)^2)$ 
and $\mu_{\rm dyn}/T_D = {\cal O}(c\cdot 10^{-3})$. 
Thus we evaluate the yield $Y_B(T_0)$ as 
\begin{align} 
Y_B(T_0) &\sim 10^{-7} \times \left(\frac{\langle \sigma(T_D/f_\pi) \rangle}{f_\pi} \right) \cdot  \left(\frac{\langle \dot{\sigma}(T_D/f_\pi) \rangle}{f_\pi^2} \right) \cdot \left( \frac{c}{1} \right)
\notag\\ 
& \sim 
 10^{-7} \times \left(\frac{\langle \sigma(T_D/f_\pi) \rangle}{f_\pi} \right) \cdot  \left(- \frac{d \langle {\sigma}(T_D/f_\pi) \rangle}{dT}\frac{\frac{T_D^2}{M_p} \cdot T_D}{f_\pi^2} \right) \cdot \left( \frac{c}{1} \right)
\notag\\ 
& \sim 
 10^{-7} \times \left(\frac{\langle \sigma(T_D/f_\pi) \rangle}{f_\pi} \right) \cdot  \left(- \frac{d \langle {\sigma}(T_D/f_\pi) \rangle}{dT}\frac{f_\pi}{M_p} \right) \cdot \left( \frac{c}{1} \right)
\notag\\ 
& \sim 
 10^{-26} \times \left(\frac{\langle \sigma(T_D/f_\pi) \rangle}{f_\pi} \right) \cdot  \left( \frac{\frac{- d \langle {\sigma}(T_D/f_\pi) \rangle}{dT} }{  {\cal O}(1)} \right) \cdot \left( \frac{c}{1} \right)
 \,. 
\end{align}
where , 
we have roughly read $d [\Delta_{l,s}(T)/\Delta_{l,s}(T=0)]/
d[T/T_{\rm pc}] |_{T\sim T_{\rm pc} \sim T_D = {\cal O}(f_\pi)} \sim - {\cal O}(1)$ 
from the lattice data on the (normalized) subtracted quark condensate $\Delta_{l,s}(T)/\Delta_{l,s}(T=0)$ versus 
$T/T_{\rm pc}$~\cite{Aoki:2009sc}. 
This thermal abundance is too small to be of the order of $10^{-10}$ observed present day. 
This is essentially due to the too slow time evolution of $\langle \dot{\sigma} \rangle$ along with the Planck scale suppression, following the adiabatic Hubble evolution.

\section{QCD with dark QCD: triggering the QCD preheating} 
\label{sec4}

In the previous section~\ref{sec3} 
we have clarified that the QCD-induced dynamic chemical potential $\mu_{\rm dyn}$ does not work for the thermal baryogenesis, due to the too slow adiabatic 
variation of $\mu_{\rm dyn}$, or equivalently, $\langle \sigma(t) \rangle$ 
in the chiral crossover transition.  
In this section we assume that QCD couples to a dQCD sector with the dark chiral 
phase transition of the first order, which makes the QCD chiral phase transition supercooled and triggers a nonadiabatic roll-down 
of $\langle \sigma(t) \rangle$.

We consider $SU(N_d)$ group for the dark color symmetry of dQCD and   
introduce one Dirac fermion $Q_{L,R}$ having 
ordinary QCD colors ($SU(3)_c$) in the fundamental representation as well: $Q_{L,R} \sim (N_d, 3)$ 
for $SU(N_d) \times SU(3)_c$. 
We set the intrinsic scale of dQCD $\Lambda_d \gtrsim \Lambda_{\rm QCD} \sim 1$ GeV.      
At the scales below $\Lambda_d \gtrsim \Lambda_{\rm QCD}$, the dark-global chiral 
$SU(3)_{dL} \times SU(3)_{dR}$ symmetry will be spontaneously 
broken by the dark quark condensate $\langle \bar{Q} Q \rangle$ down to 
$SU(3)_{dV}$.  
At almost the same moment, ordinary QCD will also be confined and 
generate the spontaneous breaking of the QCD chiral symmetry via the ordinary quark condensate $\langle \bar{q} q \rangle$. 
The low-energy description will thus form both the ordinary and dark hadron phase reflecting the double confinement as well. 

Here we give a couple of comments on the impact and sensitivity on 
the hadron phenomenology in QCD and collider physics probes to be tested in experiments today. As it turns out, the present dual QCD scenario survives 
the current terrestrial experiments.

\begin{itemize}
    \item 
At the scale $\Lambda_d$, 
the QCD color nonet $M_Q^{ab} \sim \bar{Q}_R^a Q_L^b $ 
(with $a, b$ being the QCD color indices) may couple to the QCD quarks and gluons in the gauge- and QCD chiral 
invariant way, like 
\begin{align}
    \bar{q} i \gamma^\mu D_\mu q \cdot {\rm tr}[D^\mu M_Q M_Q^\dag] 
     \,, \qquad {\rm tr} [\bar{q}_L q_R q_R q_L] \cdot {\rm tr}[M_Q^\dag M_Q] 
     \,,  \qquad 
      (\bar{q} i \gamma^\mu D_\mu q)^2 \cdot {\rm tr}[M_Q M_Q^\dag] \,,
     \qquad 
     \cdots 
    \,,   \label{portal-to-QCD}
\end{align}
where ellipses denote terms of much higher dimensional form. 
We see that the nonet couplings to quarks do not contribute to 
the minimal form of the quark-gluon gauge interaction. 
Thus the extra hadronic state will necessarily be exotic 
to be like eight-quark bound states $\sim qq qq QQ QQ$, 
protected by the dQCD and QCD color and chiral symmetries. 
Those exotic hadrons might be produced via 
the $\sigma \sigma$ scattering, $\sigma \sigma \to {\rm tr}[M_{Q}^\dag M_{Q}] \to \sigma \sigma$ including 
the portal interaction of the second type in Eq.(\ref{portal-to-QCD}), 
where the extra resonant signal would be at $\sim 2 m_{M_QM_Q} \sim 2 \Lambda_{\rm QCD} \sim 
{\rm 2 GeV}$ in the invariant mass of two $\sigma$ mesons.

\item 
The dark QCD $\eta_d$ meson $(\sim \bar{Q} i\gamma_5 Q)$ mixes with the QCD $\eta'$ meson through 
the shared axial anomaly induced from the QCD topological charge fluctuation. 
The mixing strength would be suppressed as $\sim \chi_{\rm top}/\Lambda_d^4 
\sim 10^{-4}$, where $\chi_{\rm top}$ denotes the QCD topological susceptibility, 
which is shared by $\eta_d$ and $\eta'$ and has been estimated to be $\sim (76\, {\rm MeV})^4$ as observed on the lattice~\cite{Borsanyi:2016ksw}. Thus the $\eta_d - \eta'$ mixing 
is small enough not to spoil the successful QCD $\eta'$ physics, 
as has also been discussed in the literature~\cite{Cui:2021sqx} in a different context.

\item 
Extra light quarks contribute to the running evolution of 
the ordinary QCD coupling $g_s$, where 
in the present case 
$N_d$ species of new quarks in the fundamental representation of $SU(3)_c$ group 
come into play. 
To keep the asymptotic freedom, at least $N_d$ has to be $< (33/2) -6 \sim 10$, 
which is determined by the one-loop perturbative calculation. 
Collider experiments have confirmed the asymptotic freedom with 
high accuracy 
in a wide range of higher energy scales, in particular above 10 GeV, over 1 TeV~\cite{ParticleDataGroup:2022pth}. 
When $\alpha_s$ is evolved up to higher scales 
using $\alpha_s(M_Z)$, where $\alpha_s$ is 
the fine-structure constant for $g_s$, $\alpha_s \equiv g_s^2/(4\pi)$,  
measured at the $Z$ boson pole 
($\alpha_s(M_Z)$), 
as input,  
(for a recent review of $\alpha_s$, e.g., see~\cite{Deur:2023dzc}). 
the tail of the asymptotic freedom around ${\cal O}(1)$ TeV 
can thus have sensitivity to exclude new quarks.

Current data on $\alpha_s$ at the scale around ${\cal O}(1)$ TeV  
involve large theoretical uncertainties.  
This results in uncertainty of determination of $\alpha_s(M_Z)$ for 
various experiments (LHC-ATLAS, -CMS, and Tevatron-CDF, and D0, etc.), 
which yields $\alpha_s(M_Z) \simeq 0.110 - 0.130$, being consistent with 
the world average $\alpha_s(M_Z)\simeq 0.118$ 
within the uncertainties~\cite{ParticleDataGroup:2022pth}.

We work on the two-loop perturbative computation of $\alpha_s$. 
The dQCD running coupling ($\alpha_d$) contributes to the running of $\alpha_s$ at two-loop level.  
This contribution is, however, safely negligible: 
when $\alpha_s \sim \alpha_d$ at low-energy 
scale as desired, because 
$\alpha_d \ll \alpha_s$ at high energy due to 
the smaller number of dQCD quarks (with the net number 3 coming in the beta function 
of $\alpha_d$) 
than that of the ordinary QCD quarks (with the net number 5 or 6 + $N_d$ in the beta function 
of $\alpha_s$).

Taking into account 
only the additional $N_d$ quark loop contributions to the running of $\alpha_s$, 
we thus compute the two-loop beta function, and find that 
as long as $N_d$ is moderately large $(N_d \le 5)$, 
the measured ultraviolet scaling (for 
the renormalization scale of $\mu$ = 10 GeV $-$ a few TeV) 
can be consistent with the current data~\cite{ParticleDataGroup:2022pth} within 
the range of $\alpha_s(M_z)$ above. 
Scaling down to 10 GeV, a prominent suppression  
will be seen (e.g. $\alpha_s(1\,{\rm GeV}) \sim  0.2$ for $N_{d}=5$ 
at two-loop level, compared to the perturbative-standard model prediction $\simeq 0.3$).

Precise measurements in lower scales $\lesssim 10$ GeV 
have not well been explored so far, 
due to the deep infrared complexity of QCD. 
The low-energy running of $\alpha_s$ is indeed still uncertain, and can be variant as 
discussed in a recent review, e.g., ~\cite{Deur:2023dzc}. 
The present dQCD could dramatically alter the infrared running feature of $\alpha_s$, 
due to new quarks and the running of the dQCD coupling $\alpha_d$. 
This will also supply a decisive answer to a possibility of 
the infrared-near conformality of real-life QCD.

\item 
Other stringent bounds on the extra light quarks or colored scalars 
come from the ALEPH search for gluino and squark pairs tagged with the multijets 
at the LEP experiment~\cite{ALEPH:1997mcm}. 
However, this limit has no sensitivity below the mass $\sim 2$ GeV, 
hence is not applicable to the present benchmark model.

\end{itemize}

Thus a few massless new quarks can still survive 
constraints on $\alpha_s$ at the current status~\footnote{
A similar observation has been made in~\cite{Cui:2021sqx} in a different context. 
}. 
More precise measurements of $\alpha_s$ in the future will clarify  
how many light or massless new quarks can be hidden in QCD, 
which will fix the value of $N_d$ in terms of the present dQCD.  
The present our aim to consider dQCD is mainly at the dark chiral phase transition of the first order, which is expected to be realized in a QCD-like dQCD theory with $N_d=3$ and single dark Dirac-fermion flavor ($Q$) having the ordinary QCD colors, which counts the dark flavors to be three in total. The three-flavor masless QCD-like theory has been supported to exhbit the first-order chiral phase transition, since the Pisarski-Wilczek's original work~\cite{Pisarski:1983ms} based on the perturbative renormalization group analysis and also recent functional renormalization group analysis~\cite{Fejos:2024bgl}.  
This setup is fully consistent with the current status on 
the experimental bound on the extra light quarks as above.

Now we discuss the impact of the doubly QCD charged dQCD quarks on a 
supercooled QCD chiral phase transition. 
In the double confining vacuum, 
the QCD-LSM field $M \sim q_L q_R^c$ mixes with the dQCD-QCD color singlet component   
$M_d \sim \bar{Q}_L^a Q_R^a \sim \sigma_d + i \eta_d$, 
which arises from 
a (nonperturbative) double QCD gluon exchange process in the ordinary and dark quark scattering 
with hadronization for the initial and final states taken into account: in this sense, the process would involve the mixing with 
the QCD glueball state $\sim G_{\mu\nu}^2$. 
In the double chiral invariant manner, it is generated to take the form 
\begin{align}
V_{M M_d}  = \lambda_{\rm mix} {\rm tr}
[M^\dag M] 
(M_d^\dag M_d) 
\,. \label{mix}
\end{align}
Since the generation mechanism is totally nonperturbative, 
the size of $\lambda_{\rm mix}$ is expected to be of ${\cal O}(1)$, or so. 
The dQCD sector with $N_d=3$ in total possesses 3 ($=1_{\rm dark-quark flavor} \times 3_{\rm QCD-color}$) flavors, 
so it is expected to undergo the chiral phase transition of first order and the critical temperature is $T_d = {\cal O}(100)$ MeV $(\sim T_{\rm p c})$. 
This observation is supported from 
the Pisarski-Wilczek's perturbative renormalization group argument~\cite{Pisarski:1983ms} and also recent functional renormalization group analysis~\cite{Fejos:2024bgl}. 


 \begin{figure}[t] 
\centering
	\includegraphics[scale=0.6]{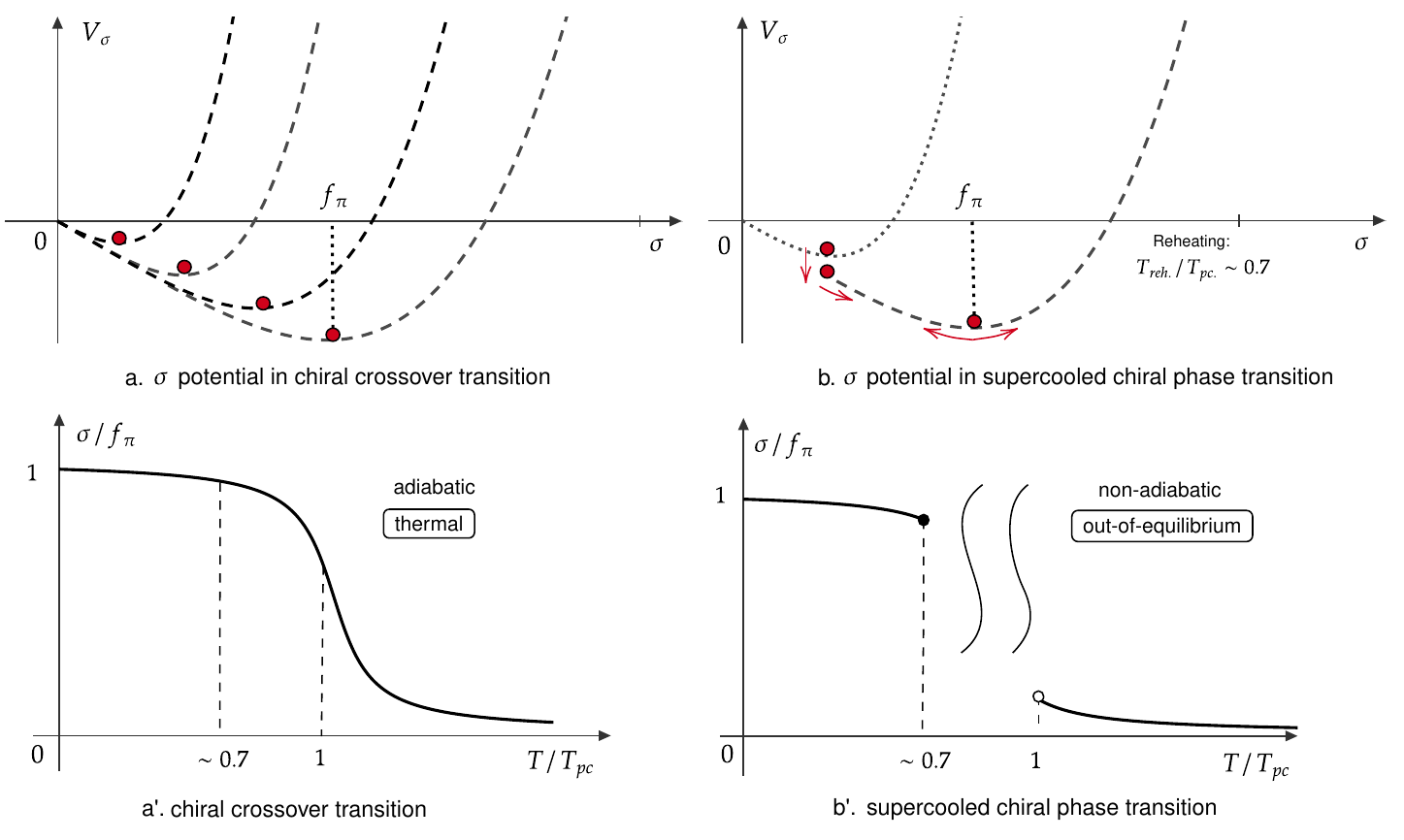}
	\caption{ The $\sigma$ potential view schematically drawing different cosmological setups between the smooth chiral crossover (panel a) and 
 supercooled chiral phase transition triggered by dQCD (panel b). In the lower panels a' and b' the corresponding thermal evolution of $\sigma$ have also been displayed. In the panels b and b', the reheating temperature is evaluated as $T_{\rm reh} \sim (0.7 - 0.8) T_{pc}$ as in the text (Eq.(\ref{Treh:value})), 
at which epoch QCD comes back to the thermal equilibrium passing through the out of equilibrium epoch illustrated by wavy curves in the center of panel b'. More detailed descriptions are discussed in the text.   
 } \label{schematic-pic}
\end{figure}

Thus at $T=T_d$, the chiral order parameter, like $\sigma_d^0$, drops down into zero. 
This is a supercooling, which stores the dark-QCD sector vacuum energy at the false vacuum. After tunneling from the false vacuum into the true vacuum, the stored vacuum energy is released into radiations (via the QCD interaction) and might slightly reheat the universe, keeping the universe in the thermal equilibrium with $T={\cal O}(T_{\rm pc})$ MeV. In the literature~\cite{Sagunski:2023ynd}, the supercooled QCD chiral phase transition, 
triggered by some dark sector contribution, 
has been addressed based on the Nambu-Jona-Lasinio description of low-energy QCD with light three flavors, where in the case of the vacuum energy as small as the QCD scale, 
the percolation temperature has been estimated to be almost identical to the 
critical temperature of the first order phase transition. 
This feature may be applicable also to the presently assumed dQCD case with three flavors (in total). Thus the latent heat generated from the dQCD chiral phase transition would be small enough that the universe can still keep almost 
the same temperature of ${\cal O}(T_{\rm pc})$.

Meanwhile, the QCD $\sigma$ gets a discontinuous drop by via the 
$\lambda_{\rm mix}$ portal coupling, which is instantaneous and much faster than 
the Hubble rate evolving along with the QCD thermal plasma. 
In the present study, we assume that the dQCD $\sigma_d$ reaches the true vacuum 
immediately after the supercooling ends, so that the QCD $\sigma$-fast roll dynamics 
solely processes decoupled from the dQCD sector. 
The $\sigma$ dropping can be observed in a drastic change of 
the QCD $\sigma$ mass term at $T=T_d$ MeV due to the induced mass squared $\Delta m^2_\sigma = \lambda_{\rm mix} f_{\pi_d}^2$ with the dark pion decay constant $f_{\pi_d} \sim \langle \sigma_d \rangle$, which, when it overwhelms the thermal mass, drives the discontinuous jump or drop of the QCD $\sigma$, and 
triggers the nonadiabatic-fast roll down around $\sigma \sim 0$. 
Including this induced $\Delta m^2_\sigma$ into the original mass square of the QCD $\sigma$, the net mass parameter realizes the usual QCD vacuum with 
$\langle \sigma \rangle = f_\pi$.  
The so-called portal coupling in Eq.(\ref{mix}) thus plays the important role to trigger the fast roll of $\sigma$ and the QCD preheating. 
Figure~\ref{schematic-pic} shows a schematic picture to describe the 
different cosmological setups between 
the chiral crossover and the dQCD-driven supercooled chiral phase transition.

\section{QCD preheating with nonadiabatic baryon chemical potential}
\label{sec5}

\begin{figure}[t] 
\includegraphics[width=0.5\linewidth]{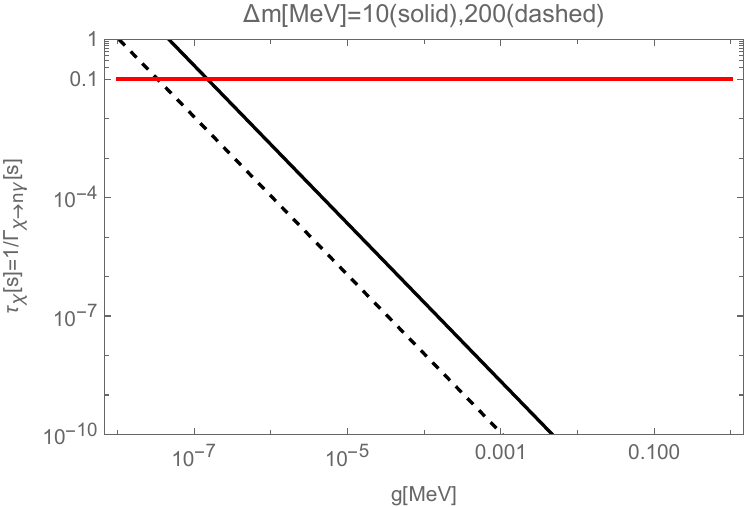} 
\caption{
The $\chi$ lifetime ($\tau_\chi$) constraints (black curves) on 
 the $n-\chi$ mixing coupling $g$ coming from 
 $\tau_\chi < \tau_{\rm BBN} \simeq 0.1$s (red straight line), given the mass difference $\Delta m$ ranged from 10 to 200 MeV. } 
 \label{lifetime-constraint}
\end{figure}

To realize the baryon number violation in the QCD chiral phase transition epoch, 
we further consider presence of another dark fermion 
fields $\chi_{L,R}$ of the SM singlet, which are 
allowed to couple to the QCD neutron fields $n_{L,R}$ in a minimal way as in the literature~\cite{Bringmann:2018sbs}: 
\begin{align}
-  {\cal L}_{n-\chi} 
  &= 
    m_D (\bar{\chi}_{R} \chi_{L})  
  + g_{L} (\bar{n}_{R} \chi_{L}) + g_{R} (\bar{\chi}_{R} n_L)+ \text{ h.c.}
\,.  \label{dark:Lag}
\end{align} 
All couplings in Eq.(\ref{dark:Lag}) are assumed to be real.  
The dark Dirac fermion $\chi$ does not carry the baryon number, so $g_{L,R}$ couplings break the baryon number $(B)$ symmetry as well as the dark fermion number ($B_\chi$) symmetry -- but, the total number of $B + B_\chi$ is conserved. 
This type of $n-\chi$ mixing coupling can be generated via leptoquark exchanges involving the $udd\chi$ vertex, in a way similar to the scenario addressed in the literature~\cite{Fornal:2018eol,Cline:2018ami,Grinstein:2018ptl,Keung:2019wpw,Keung:2020teb,Elahi:2020urr,Fornal:2020bzz,Fornal:2020gto,Fajfer:2020tqf,Alonso-Alvarez:2021oaj,Strumia:2021ybk,Berger:2023ccd,Fornal:2023wji,Liang:2023yta,Gardner:2023wyl,Bastero-Gil:2024kjo}.

The $\chi$ lifetime is constrained to escape from the BBN and/or CMB observations: when $m_\chi > m_n$ as assumed in the present study, 
$\chi$ having a sizable mixing with the neutron can be completely free from the existing cosmological and astrophysical constraints, if and only if $\chi$ decays before the BBN epoch at $t_{\rm BBN} \simeq 0.1$s or the lifetime is longer than the recombination epoch $t_{\rm rec} \sim 10^{13}$s~\cite{McKeen:2020oyr}. 
This gives the constraints on the size of the mixing couplings $g_{L,R}$. 
As in the case with the literature~\cite{McKeen:2020oyr}, 
the presently addressed $\chi$ lifetime is determined also 
by the transitioned-anomalous magnetic moment interaction:
\begin{align}
    {\cal L}_{\chi n \gamma} = \frac{\mu_n}{2} \cdot \theta \cdot \bar{\chi} \sigma_{\mu\nu} F^{\mu\nu} n + {\rm h.c.} 
    \,, \label{Lag:n-chi}
\end{align}
where $\mu_n \simeq 1.91 \mu_N$ is the neutron magnetic dipole  moment with $\mu_N = e/(2 m_n) \simeq 0.1$ e fm being 
the nuclear magneton~\cite{ParticleDataGroup:2022pth}. Here we have assumed $g_L=g_R=g$ for simplicity, and introduced the mixing angle parameter 
$\theta = g/\Delta m$ with the mass difference $\Delta m = m_\chi - m_n$. 
The $\chi \to n \gamma$ decay rate is then computed, 
for small enough $\theta$, as
\begin{align}
    \Gamma[\chi \to n \gamma] 
    \simeq \frac{1}{2200\,{\rm s}} \left( \frac{\theta}{10^{-10}} \right)^2 \Bigg| \frac{\Delta m}{10\,{\rm MeV}}  \Bigg|^3
\,. 
\end{align}
Figure~\ref{lifetime-constraint} shows the lifetime constraint on the mixing coupling $g$ for fixed $\Delta m$, which implies 
\begin{align}
    g \gtrsim 10^{-8} \, (10^{-7}) \, {\rm MeV} 
    \,, \qquad {\rm for} \qquad \Delta m = 10\, (200)\,{\rm MeV}  
    \,. \label{limit-g}
\end{align}

Combining the mixing coupling term in Eq.(\ref{Lag:n-chi}) with 
the LSM Lagrangian in Eq.(\ref{LSM-Lagrangian:eq}), 
we work on the following reduced system: 
\begin{align}
    {\cal L} &= \frac{1}{2}(\dot{\sigma})^2 
    + \bar{n} \gamma^\mu i \partial_\mu n 
    + \bar{\chi} \gamma^\mu i  \partial_\mu \chi   
    \notag\\ 
    & - \frac{m_N}{f_\pi} \sigma \bar{n}n - m_\chi \bar{\chi}\chi 
    - g (\bar{n} \chi + \bar{\chi} n) + \mu_{\rm dyn}(t) n^\dag n 
     - V_\sigma  
\,, \label{L:red}
\end{align}
where 
\begin{align}
    \mu_{\rm dyn}(t) &= \frac{c}{32 \pi^2} \frac{d}{dt} [{\sigma}^2(t)] 
\,, \notag\\ 
  V_\sigma &=  - m_\pi^2 f_\pi \sigma  + \frac{1}{2} m^2 \sigma^2  + \frac{1}{4} \lambda \sigma^4  
  \,,
\end{align}
with the same input applied to the potential parameters 
as in Eq.(\ref{LSM-Lagrangian:eq}). 
Here $\sigma$ field acts as the time-dependent space-homogeneous background field, $\sigma = \langle \sigma(t) \rangle$, 
while neutron and dark fermion fields are quantum fluctuating 
in spacetime.
At this moment the dQCD sector sigma $\sigma_d$ 
has already reached the true vacuum as noted in the previous Sec.~\ref{sec4}.

The time evolution of the vacuum expectation values 
in Eq.(\ref{net:def})
is evaluated by solving chained equations of 
motion for $n$ coupled to $\chi$ 
through Eq.(\ref{dark:Lag}), together with the dynamic $\langle \sigma (t) \rangle $: 
\begin{equation}
0=\langle\ddot{\sigma}\rangle 
+\gamma \langle\dot{\sigma}\rangle  -m_{\pi}^{2} f_{\pi}+m^{2}\langle\sigma\rangle +\lambda\langle\sigma\rangle ^{3} + \cdots 
\,. \label{EOM-sigma} 
\end{equation}
The ellipses denote the negligible terms including the Hubble friction term $(3 H \langle \dot{\sigma} \rangle )$ and the backreactions from the pion and nucleon fields~\footnote{
The QCD thermal corrections to the potential of $\sigma$ can also be safely neglected because the potential form gets almost close to the one at $T=0$, when the roll of $\sigma$ starts at around the reheating temperature $T \sim 0.7 T_{\rm pc}$, as illustrated in Fig.~\ref{schematic-pic}. This is supported from the lattice QCD result~\cite{Aoki:2009sc} which shows the subtracted light-quark condensate $\Delta_{l,s}(T) \sim \Delta_{l,s}(T=0)$ at around $T\sim 0.7 T_{\rm pc}$. 
}.
$\gamma$ plays the role of the full width of the $\sigma$ meson identified as $f_0(500)$. 
As a phenomenological input, we take $\gamma$ to be the central value of the current measurement, $\gamma \equiv |2 {\rm Im}[\sqrt{s}_{\rm pole}]| = 550$ MeV~\cite{ParticleDataGroup:2022pth}. 
Here we have applied the conventional vacuum saturation ansatz: $\langle A B\rangle 
\equiv \langle A \rangle \langle B \rangle$, which works fine in the large $N_c$ limit. 
We have also dropped the chemical potential contribution to the time evolution of $\sigma$ above, because it is quantitatively small enough due to the loop suppression associated with 
the higher-dimensional form of the operator.

Since the dynamic $\langle \sigma \rangle $ instantaneously rolls down from  $\langle\sigma\rangle \sim 0$ to $f_\pi$ with the strong friction, 
the potential energy $V(\sigma=0)-V(\sigma=f_\pi)=\frac{1}{8}(M_\sigma^2+3m_\pi^2)f_\pi^2\simeq(135\:{\rm MeV})^4$ is converted into the radiation energy. 
This triggers reheating of the universe and the reheating temperature 
can then be estimated as~\cite{Wang:2022ygk}  
\begin{align}
 T_{\rm reh} 
 &= \left[\frac{\frac{\pi^2}{30}g_*(T_{\rm reh})^4+(135\:{\rm MeV})^4}{\frac{\pi^2}{30}g_*}\right]^{1/4} 
 \notag\\ 
& \simeq 119 \: {\rm MeV} \sim 0.77 T_{pc},
\label{Treh:value}
\end{align}
where we have 
assumed the relativistic degrees of freedom $g_*(T_{\rm reh})\simeq 14$ at  $T 
= T_{\rm reh}\sim (0.7 - 0.8) T_{\rm pc}\sim 110 - 120\,{\rm MeV}$ when the dynamic $\langle \sigma \rangle $ starts to roll.  The produced thermal entropy density can also be estimated as
\begin{align} 
 \frac{2\pi^2}{45}g_*\cdot ( 120  \: {\rm MeV})^3 
  \sim 10^7 \: {\rm MeV}^3.
\label{s:estimate}
\end{align}

For practical convenience, we reformulate the fermion sector by means of the two-component spinors. 
The relations to the Dirac fermions and the definition of Dirac gamma matrices in terms of 
the two-component spinor indices are as follows:  
\begin{align}
  n&=\left(\begin{array}{c}n_L^{\alpha} \\ (n_R^{c\dagger})^{\dot{\alpha}} \end{array} \right), \qquad 
 \chi =\left(\begin{array}{c}\chi_{L}^\alpha \\ (\chi_{R}^{c\dagger})^{\dot{\alpha}} \end{array} \right) 
 \,, \notag\\ 
\gamma_\mu &= 
\left(
\begin{array}{cc}
    0 & \sigma_\mu^{\alpha \dot{\alpha}} \\
    \bar{\sigma}^{\dot{\alpha} \alpha}   & 0 
\end{array}
\right)
\,, \label{2-comp-sp}
\end{align} 
where the superscript ``$c$'' denotes the charge conjugate. 
Here  
$\alpha, \dot{\alpha}$ are spinor and dotted spinor indices acting on 
fermion and anti-fermion fields,  
and $\bar{\sigma}_i^{\dot{\alpha} \alpha}= - \sigma_i^{\dot{\alpha}\alpha}$ for $i=1,2,3$, 
$\sigma_0^{\alpha \dot{\alpha}}=1^{\alpha \dot{\alpha}}$ and $\bar{\sigma}_0^{\dot{\alpha}\alpha}=1^{\dot{\alpha} \alpha}$. 
Then Eq.(\ref{L:red}) is cast into the form 
\begin{align}
    {\cal L} &= \frac{1}{2}(\dot{\sigma})^2 + 
    n^\dag_L i \bar{\sigma}^\mu \partial_\mu n_L 
    + n^{c}_R i \sigma^\mu \partial_\mu n_{R}^{ c \dag} 
    +  \chi^\dag i \bar{\sigma}^\mu \partial_\mu \chi_L 
    + \chi^{c}_R i \sigma^\mu \partial_\mu \chi_{R}^{c \dag}  
    \notag\\ 
    & -\frac{m_N}{f_\pi} \sigma ({n}^\dag_L n_R^{c \dag} + {\rm h.c.})  
     - m_\chi ({\chi}^\dag \chi_R^{c \dag} + {\rm h.c.})  
    - g (  n_L \chi_R^{c} + 
    \chi_L n_R^{c} + {\rm h.c.} ) 
    \notag\\ 
    & + \mu_{\rm dyn}(t)\left(
n_{L}^{\dagger} \bar{\sigma}_0 n_{L} + n_{R}^{c} {\sigma}_0 n_{R}^{c \dag}  \right) 
    \notag\\ 
    & - V  
\,.  \label{L:red2}
\end{align} 
The relevant equations of motion for the fluctuating spinor fields are: 
\begin{equation}
\begin{aligned}
0 &=\bar{\sigma}^{\mu} i \partial_{\mu} n_{L}+ \tilde{m}_N  n_{R}^{c \dagger} 
+ g \chi_{R}^{c \dagger} 
+ \mu_{\rm dyn}(t) \bar{\sigma}_0 n_L   
\,,  \\
0 &=\sigma^{\mu} i \partial_{\mu} n_{R}^{c \dag}  
+ \tilde{m}_{N} n_{L}
+ g  \chi_{L} 
+  \mu_{\rm dyn}(t) \sigma_0 n_R^{c \dag} 
\,,  \\
0 &=\bar{\sigma}^{\mu} i \partial_{\mu} \chi_{L} + m_\chi \chi_{R}^{c \dagger} 
 + g  n_{R}^{c \dagger} \,,  \\
0 &=\sigma^{\mu} i \partial_{\mu} \chi_{R}^{c \dag} 
+ m_\chi \chi_{L} 
+ g n_{L}\,, 
\label{EOMs:spinors}
\end{aligned}
\end{equation}
where 
\begin{align}
  \tilde{m}_N(t)=m_N\frac{\langle\sigma(t)\rangle }{f_\pi}
\, . \label{mNtilde} 
\end{align} 
The time-varying mass $\tilde{m}_N$ in Eq.(\ref{mNtilde}) causes the nonperturbative nucleon production when the adiabaticity is violated: $|\dot{\tilde{m}}_N/\tilde{m}_N^2|\gtrsim 1$.  This inequality leads to the production range in terms of the $\sigma$ motion as 
\begin{equation}
 |\sigma| \lesssim \sqrt{\frac{f_\pi \langle\dot{\sigma}\rangle }{m_N}}\simeq 42 \:{\rm MeV}, \label{production_area:eq}
\end{equation}
where we read $\langle \dot{\sigma}\rangle \simeq \sqrt{V(\sigma=0)-V(\sigma=f_\pi)}\sim (135\:{\rm MeV})^2$.
The actual production time can be earlier because the estimated velocity $\langle\dot{\sigma}\rangle $ would be smaller due to the friction $\gamma$.

We move on to the momentum space by the Fourier transform: 
\begin{equation}
\left(\begin{array}{c}
n_{L}(t, \vec{x}) \\
n_{R}^{cT}(t, \vec{x}) \\
\chi_L(t, \vec{x}) \\
\chi_{R}^{cT}(t, \vec{x})
\end{array}\right)_{\alpha}=\int \frac{d^{3} k}{(2 \pi)^{3}} e^{i \vec{k} \cdot \vec{x}} \sum_{s=\pm}\left(e_{\vec{k}}^{s}\right)_{\alpha}\left(\begin{array}{c}
n_{L, \vec{k}}^{s}(t) \\
n_{R, \vec{k}}^{s}(t) \\
D_{L, \vec{k}}^{s}(t) \\
D_{R, \vec{k}}^{s}(t)
\end{array}\right)
\,,  \label{FT}
\end{equation}
where we have introduced the helicity eigenstate vector having helicity $s=+,-$ as 
\begin{equation}
\left(e_{\vec{k}}^{s}\right)_{\alpha} = 
\left( 
\begin{array}{c} 
  \sqrt{\frac{1}{2}  \left(1+\frac{s k^{3}}{k}\right)}  \\   
   s \, e^{i \theta_{k}} \sqrt{\frac{1}{2}\left(1-\frac{s k^{3}}{k}\right)}
\end{array}
\right) 
\,,
\end{equation}
with $k \equiv |\vec{k}|$ and  
\begin{equation}
e^{i \theta_{\vec{k}}} \equiv \frac{k^{1}+i k^{2}}{\sqrt{\left(k^{1}\right)^{2}+\left(k^{2}\right)^{2}}}
\,,
\end{equation}
which are the solutions to the helicity-eigen equations for 
left- and right-handed spinors: 
\begin{align}
    - \frac{k^i}{k} (\sigma_0 \bar{\sigma}_i)^\beta_\alpha \cdot (e_{\vec{k}}^s)_\beta 
    &= s \cdot (e^s_{\vec{k}})_\alpha 
    \,, \notag \\ 
- \frac{k^i}{k} (\bar{\sigma}_0 \sigma_i)_{\dot{\beta}}^{\dot{\alpha}} 
\cdot (e_{\vec{k}}^{s \dag})^{\dot{\beta}}  
    &= s \cdot (e^{s \dag}_{\vec{k}})^{\dot{\alpha}}  
    \,.  
\end{align} 
Applying those transformed fields into equations of motion in Eq.(\ref{EOMs:spinors}), 
  we get 
\begin{equation}
\begin{aligned}
{\cal D}_{t} n_{L, \vec{k}}^{s} &=i s k n_{L, \vec{k}}^{s}+i s e^{-i \theta_{\vec{k}}}\left(\tilde{m}_{N} n_{R,-\vec{k}}^{s \dagger}+g  D_{R,-\vec{k}}^{s \dagger}  \right) 
\,,  \\
{\cal D}_{t} n_{R, \vec{k}}^{s} &=i s k n_{R, \vec{k}}^{s}+i s e^{-i \theta_{\vec{k}}}\left(\tilde{m}_{N} n_{L,-\vec{k}}^{s \dagger}+g D_{L,-\vec{k}}^{s \dagger}\right) 
\,,  \\
{\cal D}_{t} D_{L, \vec{k}}^{s} &=i s k D_{L, \vec{k}}^{s}+i s e^{-i \theta_{\vec{k}}}\left(m_{\chi} D_{R,-\vec{k}}^{s \dagger} +g n_{R,-\vec{k}}^{s \dagger}\right)\,,  \\
{\cal D}_{t} D_{R, \vec{k}}^{s} &=i s k D_{R, \vec{k}}^{s}+i s e^{-i \theta_{\vec{k}}}\left(m_{\chi} D_{L,-\vec{k}}^{s \dagger} +g n_{L,-\vec{k}}^{s \dagger}\right)\,, 
\end{aligned}
\label{EOMs}
\end{equation} 
where 
\begin{align}
    {\cal D}_{t} n_{L, \vec{k}}^{s} &= (\partial_t  - i \mu_{\rm dyn}(t)) n_{L, \vec{k}}^{s}
\,, \notag\\ 
 {\cal D}_{t} n_{R, \vec{k}}^{s} &= (\partial_t  + i \mu_{\rm dyn}(t)) n_{R, \vec{k}}^{s}
\,, \notag\\ 
{\cal D}_{t} D_{L, \vec{k}}^{s} &= \partial_t D_{L, \vec{k}}^{s}
\,, \notag\\ 
{\cal D}_{t} D_{R, \vec{k}}^{s} &= \partial_t D_{R, \vec{k}}^{s}
\,, 
\end{align}
and we have used the orthogonality and completeness relations for helicity eigenstates, 
\begin{align}
    & e^{s \dag}_{\vec{k}} \bar{\sigma}_0 e^{s'}_{\vec{k}} = 
    e^{s}_{\vec{k}} {\sigma}_0 e^{s' \dag}_{\vec{k}} = \delta^{ss'} 
    \,, \notag\\ 
    & \sum{s,s'} (e^s_{\vec{k}})_\alpha (e^{s \dag}_{\vec{k}})_{\dot{\alpha}} 
    = (\sigma_0)_{\alpha \dot{\alpha}} 
    \,, \qquad 
    \sum{s,s'} (e^{s \dag}_{\vec{k}})^{\dot{\alpha}} (e^{s}_{\vec{k}})^{{\alpha}} 
    = (\bar{\sigma}_0)_{\dot{\alpha} \alpha} 
    \,, \notag\\ 
    & e^{s}_{\vec{k}} e^{s'}_{-\vec{k}} = s \, e^{i \theta_{\vec{k}}} \cdot \delta^{ss'} \,, \qquad 
    e^{s \dag}_{-\vec{k}} e^{s' \dag}_{\vec{k}} = s \, e^{-i \theta_{\vec{k}}} \cdot \delta^{ss'} 
    \,. 
\end{align} 
Because of the spacial-homogeneous universe we are allowed to focus only on one mode ($n_k$) having $\vec{k}=\left(\epsilon, 0, k_{3}\right)$ with $\epsilon \rightarrow 0$ in the end, 
which leads to $e^{i \theta_{\vec{k}}}=1$, 
and then perform the momentum-space integral $\int 4\pi  k^2 n_k dk$.

The net baryon number density is computed as 
    \begin{align}
n_{B}(t)
&= \frac{1}{V} \int d^{3} x 
\left(
\langle 
n^\dag  n 
\rangle 
-\sum_{\vec{k}}2\right)
\,, \label{net:def}
\end{align}
where $V=\int d^{3}x$ is the space volume, and 
the last term corresponds to the subtraction of the divergent part induced by the zero-point energy ($4\times \frac{1}{2}$). 
The net baryon number density in Eq.(\ref{net:def}) 
can be rewritten in terms of the vacuum expectation values of two Fourier transformed fields as 
\begin{equation}
\begin{aligned}
n_{B}(t)  
&=\frac{1}{V} \int \frac{d^{3} k}{(2 \pi)^{3}} \sum_{s=\pm}\left(\left\langle n_{L, \vec{k}}^{s \dagger} n_{L, \vec{k}}^{s}\right\rangle -\left\langle n_{R, \vec{k}}^{s \dagger} n_{R, \vec{k}}^{s}\right\rangle \right) 
\label{nB}
\end{aligned}
\,, 
\end{equation} 
The two-point functions relevant to this net baryon number density 
are developed in time through the time evolution equations given in Eq.(\ref{EOMs}). 
The explicit expressions of those time evolution equations and 
the initial conditions for the two-point functions are supplemented in 
Appendix~\ref{two-point}.

\section{Baryogenesis via QCD preheating}
\label{sec6}

In this section, we numerically evaluate the time-evolution of $\langle \sigma(t) \rangle$, the chemical potential $\mu_{dyn}(t)$, and the BNA $Y_B=\frac{n_B}{s}$, based on the formulae presented in the previous Sec.~\ref{sec5} and listed in Appendix~\ref{two-point}. 
The entropy density $s$ is set as the thermal one, in Eq.(\ref{s:estimate}), 
which turns out to be comparable with the nonperturbative pionic entropy density 
created through the preheating as noted in~\cite{Wang:2022ygk}.

\begin{figure}[t]
    \centering
    \includegraphics[width=0.65\linewidth]{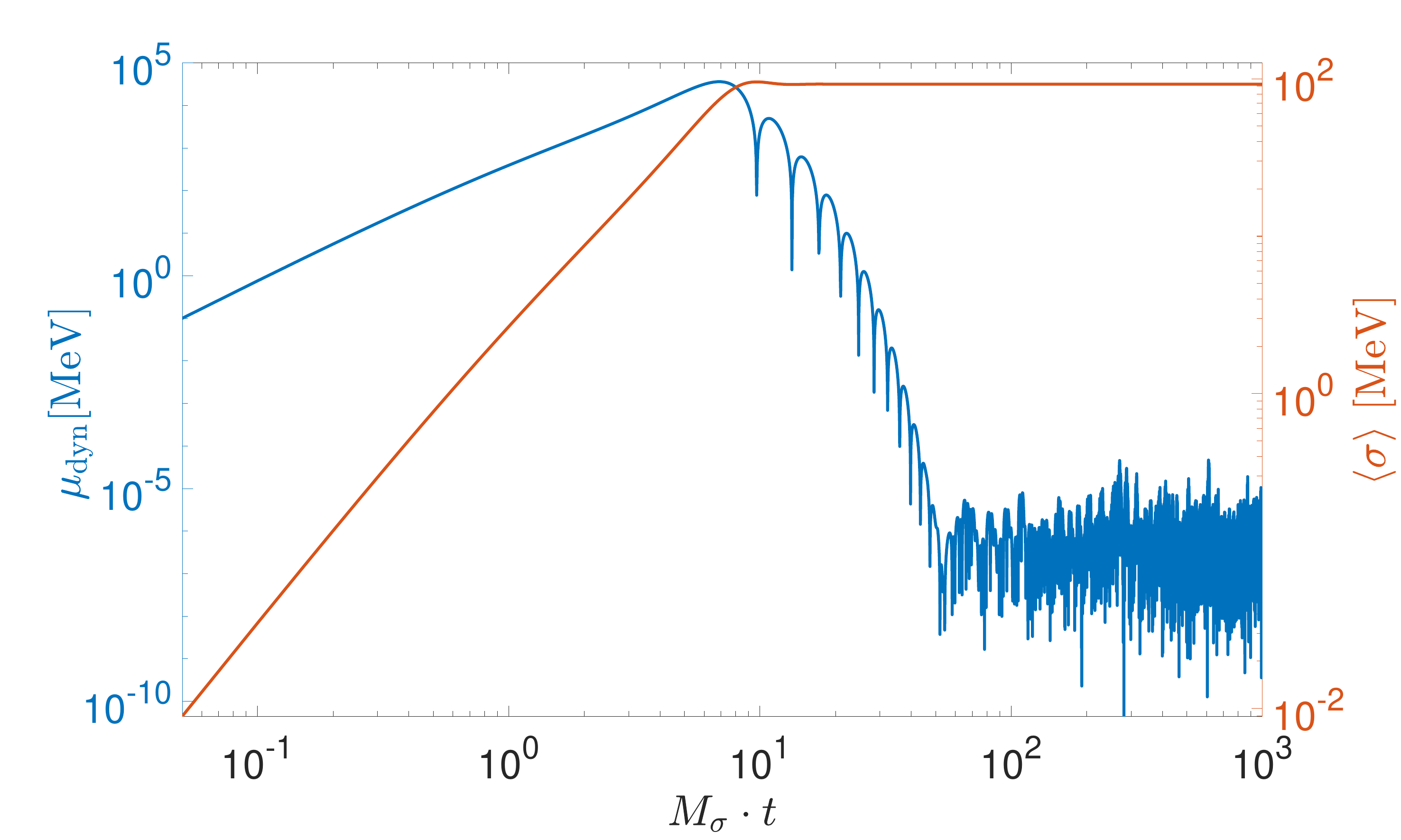}
    \caption{The time evolution of $\langle \sigma (t) \rangle$ (orange curve) 
    and $\mu_{\rm dyn}$ in Eq.(\ref{mudyn}) (blue curve) in the QCD preheating epoch defined in the time interval $0 \le M_\sigma t \le 700$ as in Eq.(\ref{relax:time}). } 
    \label{sigma-mu}
\end{figure}

$\left\langle \sigma(t) \right\rangle$ is 
evaluated by solely solving the space-homogeneous (approximate) time-evolution Eq.(\ref{EOM-sigma}) with the input parameters setup as in the previous sections. 
First of all, we recall that 
the preheating scalar $\sigma$ starts to roll down from $\sigma \sim 0$, which is driven 
via the portal coupling in Eq.(\ref{mix}) 
instantaneously just after the dQCD $\sigma_d$ undergoes the first-order phase 
transition at $T\sim f_{\pi_d}= {\cal O}(100)$ MeV, as in the right panel of 
Fig.~\ref{schematic-pic}. 
Thus we can simply set the initial condition of $\langle \sigma(t) \rangle$ as 
\begin{align}
    \langle \sigma(t=t_0) \rangle = \langle \dot{\sigma}(t=t_0) \rangle = 0 
\,. 
\end{align} 
The nonzero $\langle \sigma(t=0) \rangle$ will not substantially alter 
the essential features of the nonadiabatic evolution of $\langle \sigma(t) \rangle$. 
Recall that the nonadiabatic condition is met when 
$|\sigma| \lesssim \sqrt{\frac{f_\pi \langle\dot{\sigma}\rangle }{m_N}}\simeq 42 \:{\rm MeV}$ (Eq.(\ref{production_area:eq})), which corresponds to  
\begin{equation}
    M_{\sigma} \cdot t \lesssim 5
    \,. \label{prod:range:2}
\end{equation}
See the orange curve in Fig.~\ref{sigma-mu}. 
Since the time scale of the $\sigma$ motion ($\sim M_\sigma^{-1}$) is thus much shorter than the Hubble rate $(\sim M_p/T^2_{\rm pc})$, the produced neutrons and 
pions will never be thermalized. 

The total nucleon number density can be evaluated as 
\begin{eqnarray}
 n_N(t) + \bar{n}_N(t)
  &=& \sum_{\vec{k}}\frac{\langle \tilde{\rho}_N(\vec{k},t)\rangle }{\omega(\vec{k},t)} 
  \simeq 2 \sum_{\vec{k}}\frac{\langle \tilde{\rho}_n(\vec{k},t)\rangle }{\omega(\vec{k},t)} 
  \,.  \label{total_baryon} 
\end{eqnarray}
for $N=p,n$, where $\vec{A}$ denotes the spatial component of the vector variable $A$ and $\omega(\vec{k},t)=\sqrt{|\vec{k}|^2+\tilde{m}_N^2(t)}$ is the  one-particle energy of the nucleon.   
In reaching the second equality of Eq.(\ref{total_baryon}) we have simply taken the isospin symmetric limit. 
$\tilde{\rho}_n(\vec{k},t)$ is the kinetic energy density 
of the nucleon in momentum space, which is derived from the Hamiltonian as~\footnote{
 A similar formula for the energy density in~\cite{Wang:2022ygk} includes errors, which have fully been corrected in Eq.(\ref{total_baryon}). } 
\begin{align} 
 \langle \tilde{\rho}_n(\vec{k},t) \rangle 
  &=   
  \frac{1}{V}   \sum_{s= \pm}\left[\omega(\vec{k},t) - s|\vec{k}| \left(\left\langle n_{L, \vec{k}}^{s \dagger} n_{L, \vec{k}}^s\right\rangle+\left\langle n_{R, \vec{k}}^{s \dagger} n_{R, \vec{k}}^s\right\rangle\right)\right. \notag \\
& \left. + s \, \tilde{m}_N \left( e^{i \theta_{\vec{k}}}\left\langle n_{R,-\vec{k}}^s n_{L, \vec{k}}^s\right\rangle+ e^{-i \theta_{\vec{k}}}\left\langle n_{L, \vec{k}}^{s \dagger} n_{R,-\vec{k}}^{s \dagger}\right\rangle\right)\right]
  \,. 
  \label{energy_density} 
\end{align}
The detailed derivation of Eq.(\ref{energy_density}) is  supplemented in Appendix~\ref{BT}.  
The last term in  Eq.(\ref{energy_density}) corresponds to the subtraction of the negative vacuum energy ($4\times \frac{1}{2}\omega(\vec{k},t)$).
The time evolution of the total baryon number density 
in Eq.(\ref{total_baryon}) is thus determined by solving coupled equations of 
motion for $n_{L, \vec{k}}^s$, $n_{R, \vec{k}}^s$, and $\langle \sigma(t) \rangle $ 
(Eqs.(\ref{EOMs}) and (\ref{EOM-sigma})). 
Note that as long as the $n-\chi$ mixing coupling $g$ is small enough compared to the typical scale of the preheating frequency $\sim {\cal O}(M_\sigma)$, the total number density $(n_N + \bar{n}_N)$ is almost completely determined by QCD without dQCD, as was stressed in~\cite{Wang:2022ygk}. The dynamical chemical potential term has been also dropped in there, because the total number density $(n_N + \bar{n}_N)$ is dominantly produced via the strong Yukawa coupling $\sim m_N/f_\pi={\cal O}(10)$.

$(n_N + \bar{n}_N)$ is explosively generated by the nonadiabatic-time varying $\langle \sigma (t) \rangle $
and gets asymptotically saturated to be $ \simeq 10^5 \ {\rm MeV}^3$
in the time range $M_\sigma t\lesssim 5$ (Eq.(\ref{prod:range:2})). 
This number density 
is much larger than the thermal equilibrium density at the reheating temperature $T_{\rm reh}\simeq  119$ MeV (Eq.(\ref{Treh:value})), 
$[(n_N + \bar{n}_N)]_{\rm EQ} = 2 \times 2 \times 2 \times (m_N T_{\rm reh}/2\pi)^{3/2} e^{-m_N/T_{\rm reh}} \sim  7 \times 10^3 \ {\rm MeV}^3$, 
hence becomes overabundant. 
Actually, the overproduced nucleons can survive long enough during the whole reheating process: 
the relaxation time scale, at which the overproduced nucleons pair-annihilate,  
can be estimated as 
\begin{align} 
M_\sigma \Delta t_{\rm relax}=M_\sigma(n_N \langle \sigma v \rangle )^{-1}\sim 700 
\, , \label{relax:time}
\end{align}  
or equivalently,  $\Delta t_{\rm relax} \sim 5 \times 10^{-7} \ {\rm fs} $ for $M_{\sigma = f_0(500)}=500$ MeV. 
Here the static nucleon-pair annihilation 
cross section has been evaluated as a classical disc 
$\langle \sigma v \rangle  \sim 4\pi/m_N^2$ 
with the impact parameter $(1/m_N)$.  
Thus the QCD preheating is operative to accumulate a large number of nucleon and antinucleon pairs in out-of-equilibrium until the relaxation time. 
This is a novel picture of the thermal history 
in the QCD phase transition epoch, as emphasized in the literature~\cite{Wang:2022ygk}.

\begin{figure}[t]
    \centering
    \includegraphics[width=0.65\linewidth]{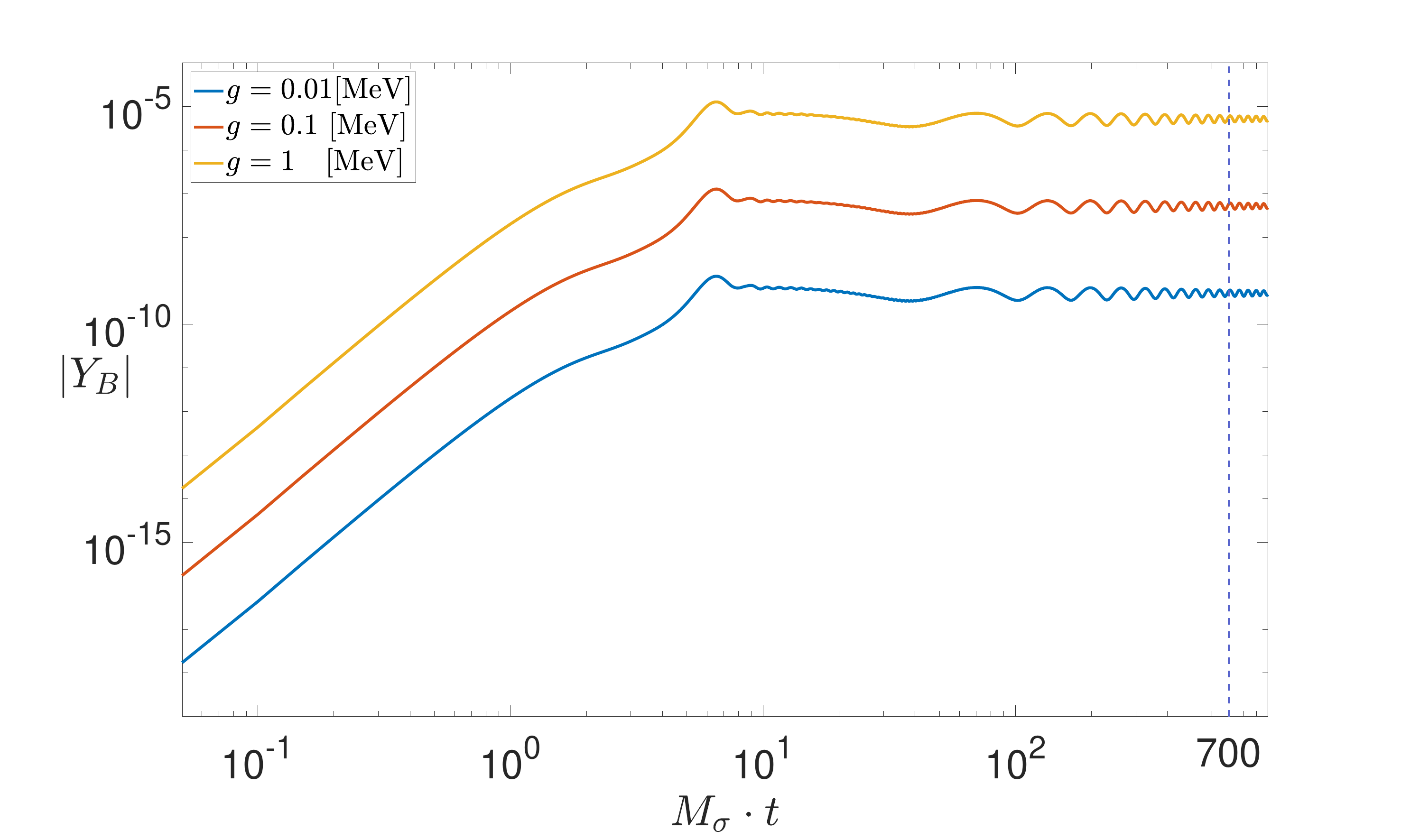}
    \caption{The time evolution and freeze out of BNA in the QCD preheating epoch, as in Fig.~\ref{sigma-mu}, for various sizes of the $n-\chi$ mixing coupling strength $g$. The dashed vertical line at $M_\sigma t = 700$ corresponds to the relaxation time scale in Eq.(\ref{relax:time}) until which scale the QCD preheating needs to be finalized. In this figure we have taken $m_N = 1000$ MeV, $m_\chi=1050$ MeV, and $c=1$, as a reference point. For the parameter dependence, see the text.}  
    \label{BNA-g}
\end{figure}

In Fig~\ref{sigma-mu} we show the time evolution of 
$\langle \sigma(t) \rangle$ (orange curve) 
and $\mu_{\rm dyn}(t)$ (blue curve) until around 
the relaxation time scale in Eq.(\ref{relax:time}). 
$\langle \sigma(t) \rangle$ reaches the true vacuum $\langle \sigma \rangle = f_\pi $ at 
$M_\sigma t \sim 10$ passing through the nonadiabatic particle production regime $M_\sigma t < 5$ in Eq.(\ref{prod:range:2}). 
It then goes through the true vacuum $\langle \sigma \rangle= f_\pi$ to be pulled back and trapped at the true vacuum due to 
the strong friction of the $\sigma$ decay width $\gamma$. 
Thus the $\langle \sigma (t)\rangle$ evolution does not simply go like $\sim \cos M_\sigma t$, hence the particle production does not follow the Mathieu equation or the parametric resonance~\cite{Wang:2022ygk}.

The dynamic baryon chemical potential $\mu_{\rm dyn}$, 
which essentially scales with $\langle \dot{\sigma} \rangle$ as in Eq.(\ref{mudyn}), 
grows up until $\langle \sigma(t) \rangle$ reaches the true vacuum, passing through the nonadiabatic production epoch. 
It then drastically damps due to $\gamma$, and finally freezes out 
at $M_\sigma t = {\cal O}(100)$ (which keeps exhibiting a tiny intrinsic oscillation in magnitude, though).

Figure~\ref{BNA-g} displays the BNA evolved until the relaxation time scale $M_\sigma t \sim 700$ with varying the $n-\chi$ mixing strength $g$ (in unit of MeV). 
The BNA yield $Y_B$ grows in magnitude until $\langle \sigma(t) \rangle$ reaches the true vacuum for the first time (at $M_\sigma t \sim 10$ as seen in Fig.~\ref{sigma-mu}). 
The BNA yield $Y_B$ is essentially produced by the nonadiabatic $\sigma$ motion   
and the associated nonperturbative nucleon-antinucleon pair production
for the time scale in the production era  
$0 \le M_\sigma t \lesssim 5$.  
Finally,  $Y_B$ asymptotically gets saturated to constant 
due to the decoupling of $\mu_{\rm dyn}$, 
which starts to operate at around $M_\sigma t \sim 10$ (see Fig.~\ref{sigma-mu}) 
and drives the $n -\bar{n}$ oscillation to end until the relaxation time scale 
$M_\sigma t \sim 700$.

The size of the net baryon number density $n_B$ evolves like $|n_B(t)| \sim  | \frac{\mu_{\rm dyn}(t)  g^4}{m_\chi^2 - m_N^2}|$, 
so simply grows with the $n-\chi$ mixing coupling strength $g$ and the size of 
the dynamic baryon chemical potential, $c$ in Eq.(\ref{mudyn}). 
We have also checked that 
the dependence of the dark fermion mass $m_\chi$ is small 
enough that the BNA time evolution until the relaxation time scale 
is not substantially altered from the case with $m_\chi=1050$ MeV, as in Fig.~\ref{BNA-g}, as long as $m_\chi$ is close, but greater than $m_N$ (such as in a range of $1000 < m_\chi < 1200$ MeV).

The baryogenesis is completed before the relaxation time period ($M_\sigma t \sim 700$) and, with $g = {\cal O}(10^{-2})$ MeV, successfully yield the desired amount of the baryon asymmetry, $Y_B = n_B/s \sim 10^{-10} - 10^{-9}$, 
consistently with the current cosmological and astrophysical constraint in Eq.(\ref{limit-g}). 
Thus, the nonadiabatic $\mu_{\rm dyn}(t)$ significantly works for the 
baryogenesis via the QCD preheating.

\section{Summary and Discussions} 
\label{sec:sumamry}

In summary, 
the dynamic chemical potential induced from QCD can provide the key role to 
generate the BNA in a nonadiabatic way through the QCD preheating played 
by the light quark condensate. The idea of the generation mechanism with the induced chemical potential is analogous to the original spontaneous 
baryogenesis~\cite{Cohen:1987vi}, which was based on the adiabatic (thermal) generation, and the leptogenesis via the Higgs relaxation or inflation~\cite{Kusenko:2014lra,Yang:2015ida,Pearce:2015nga,Gertov:2016uzs,Kawasaki:2017ycl,Wu:2019ohx,Lee:2020yaj,Cado:2021bia}, the axion(like) scalar oscillation~\cite{Kusenko:2014uta,Ibe:2015nfa,Daido:2015gqa,Takahashi:2015waa,Adshead:2015jza,Takahashi:2015ula,Kusenko:2016vcq,Maleknejad:2016dci,DeSimone:2016ofp,DeSimone:2016juo,Son:2018avk,Dasgupta:2018eha,Bae:2018mlv,Domcke:2019qmm,Wu:2020pej,Berbig:2023uzs}.  
In contrast to the latter cases, the presently addressed scenario 
does not assume anything at higher scales beyond the SM for generating the chemical potential operator, which arises from QCD of the SM, instead, 
predicts new physics sector (what we call dQCD and a dark Dirac fermion $\chi$) exhibiting the cosmological phase transition with the supercooling at around the QCD scale.

The presently proposed baryogenesis includes two dark sectors: a dark fermion $\chi$ and dQCD with QCD colors. In closing, we give several comments on the phenomenological and cosmological probes related to those dark sectors, 
in addition to the impacts on the QCD hadron physics and TeV scale probes at the LHC as noted in Sec.IV.

The dark fermion $\chi$ coupled to the neutron, which violates the baryon number symmetry, can be present with the mass of $\sim 1$ GeV consistently with 
the current cosmological and astrophysical bound (Eq.(\ref{limit-g})) 
and the desired BNA yield (Fig.~\ref{BNA-g}). 
This baryon number violating coupling may imply another new physics at higher scales, 
like leptoquarks as addressed in the literature~\cite{Fornal:2018eol,Cline:2018ami,Grinstein:2018ptl,Keung:2019wpw,Keung:2020teb,Elahi:2020urr,Fornal:2020bzz,Fornal:2020gto,Fajfer:2020tqf,Alonso-Alvarez:2021oaj,Strumia:2021ybk,Berger:2023ccd,Fornal:2023wji,Liang:2023yta,Gardner:2023wyl,Bastero-Gil:2024kjo},   
which would be constrained by the current experimental limits on searching for leptoquarks (for a review, e.g., see~\cite{Dorsner:2016wpm}). 
Thus the allowed baryon number violating coupling strength $g$ could be sensitive to those leptoquark searches, depending on the modeling, which might be challenging to generate the desired BNA as in Fig.~\ref{BNA-g} and to even survive the BBN limit in Eq.(\ref{limit-g}).

The leptoquark physics would also discriminate the present scenario 
and the original QCD preheating setup~\cite{Wang:2022ygk}, in which $\mu_{\rm dyn}(t)$ was simply assumed to be negligible, instead, both the CP violation and the source to freeze out the BNA 
were provided from another dark sector. In particular, in the original work~\cite{Wang:2022ygk} an axionlike particle having 
a time-damping evolution has been assumed to make the baryon number violating coupling dropped to zero, so as to end the $n -\bar{n}$ oscillation. 
Searching for leptoquarks as well as the axionlike particle could thus be 
a punchline to test whether the QCD-induced dynamic baryonic chemical potential can be sizable and baryon number violating coupling is still present today.

The supercooled dQCD in the present scenario can generate the gravitational wave spectra associated with the percolation/nucleation temperature around the QCD scale. The signals could be to be probed via the pulsar timing arrays sensitive to the peak frequencies around nano hertz~\cite{NANOGrav:2023gor,NANOGrav:2023hfp,EPTA:2023sfo,EPTA:2023akd,EPTA:2023fyk,Reardon:2023gzh,Reardon:2023zen,Xu:2023wog}, or the future prospected interferometers sensitive to higher frequencies~\cite{LISA:2017pwj,Caprini:2019egz,Corbin:2005ny,Harry:2006fi,Kawamura:2006up,Yagi:2011wg}. 
Consistency with the realization of the QCD preheating setup, 
as illustrated in Fig.~\ref{schematic-pic}, can thus be discussed in light of the gravitation wave prediction.

Dark matter candidates can arise from the dQCD sector, such as 
the dark baryons and/or mesons. The sizable enough relic abundance could be generated through another preheating which can arise 
due to the nonadiabatic oscillation of the dark sigma meson $\sigma_d$, triggered by the dQCD supercooling, right before the QCD preheating epoch. 
Since the $\sigma- \sigma_d$ portal coupling is present and 
the dQCD as well as QCD interactions are strong, 
the dark-matter direct detection experiments could probe those dQCD dark-matter hadrons.

Thus, the presently proposed and explored baryogenesis 
can involve plenty of phenomenological and cosmological probes 
to be tested by the current and future prospected observations 
in the sky and/or terrestrial experiments. 
Those would be deserved to purse in the future publications.

\section*{Acknowledgments} 

We are grateful to Seishi Enomoto, Yuepeng Guan, Hiroyuki Ishida, and Jinyang Li for fruitful  discussions. 
This work was supported in part by the National Science Foundation of China (NSFC) under Grant No.11747308, 11975108, 12047569, 
and the Seeds Funding of Jilin University (S.M.).

\appendix

\section{Derivation of Eq.(\ref{energy_density}) } 
\label{BT}

\subsection{Derivation of Eq.(\ref{energy_density})}

We consider the relevant part of the Lagrangian in Eq.(\ref{L:red2}):
\begin{equation}
     {\cal L}= 
     n_L^\dag i \bar{\sigma}^\mu \partial_\mu n_L 
    + n^{c}_R i \sigma^\mu \partial_\mu n_{R}^{ c \dag} 
    - \tilde{m}_N({n}_L^\dag n_R^{c \dag} + {\rm h.c.})  
\,. 
\end{equation}
From this we read off the canonical conjugate fields, 
\begin{equation}
\tilde{n}_L
=\frac{\partial{\cal L}}{\partial \dot{n}_L}
=n^{\dag}_L i \bar{\sigma}_0; \ \ \ \ 
\tilde{n}^{c \dag}_R
=\frac{\partial{\cal L}}{\partial \dot{n}^{c \dag}_R}
=n^{c}_R i \sigma_0
\,. 
\end{equation}
By performing the Legendre transformation, we get the Hamiltonian
\begin{equation}
\begin{aligned}
H=&\int{d^3x}(
\tilde{n}_L \dot{n}_L + \tilde{n}^{c \dag}_R \dot{n}^{c \dag}_R
-{\cal L})\\
=&\int{d^3x}[
- n_L^{\dag} i \bar{\sigma}^i \partial_i n_L
- n_R^{c} i \sigma^i \partial_i n_R^{c \dag}
+\tilde{m}_N(
n_L^{\dag}n_R^{c \dag}+h.c.
) ]\,. 
\end{aligned}
\end{equation}
This can further be evaluated in terms of the Fourier transformed fields in Eq.(\ref{FT}), as follows: 
\begin{equation}
\begin{aligned}
H=\int{d^3x}&[
-\int{\frac{d^3k}{(2\pi)^3}} e^{-i \vec{k} \cdot \vec{x}}
\sum_{s} e_{\vec{k}}^{s\dag} n_{L,\vec{k}}^{s\dag}
\cdot i \bar{\sigma}^i 
\int{\frac{d^3k^{\prime}}{(2\pi)^3}} (i\vec{k}^{\prime})
e^{-i \vec{k}^{\prime} \cdot \vec{x}}
\sum_{s^{\prime}} e_{\vec{k}^{\prime}}^{s^{\prime}}
n_{L,\vec{k}^{\prime}}^{s^{\prime}}\\
&-\int{\frac{d^3k}{(2\pi)^3}} e^{i \vec{k} \cdot \vec{x}}
\sum_{s} e_{\vec{k}}^{s} n_{R,\vec{k}}^{s}
\cdot i \sigma^i 
\int{\frac{d^3k^{\prime}}{(2\pi)^3}}(-i\vec{k}^{\prime})
e^{-i \vec{k}^{\prime} \cdot \vec{x}}
\sum_{s^{\prime}} e_{\vec{k}^{\prime}}^{s^{\prime}\dag}
n_{R,\vec{k}^{\prime}}^{s^{\prime}\dag}\\
&+\tilde{m}_N (
\int{\frac{d^3k}{(2\pi)^3}} e^{-i \vec{k} \cdot \vec{x}}
\sum_{s} e_{\vec{k}}^{s\dag} n_{L,\vec{k}}^{s\dag}
\int{\frac{d^3k^{\prime}}{(2\pi)^3}} 
e^{-i \vec{k}^{\prime} \cdot \vec{x}}
\sum_{s^{\prime}} e_{\vec{k}^{\prime}}^{s^{\prime}\dag}
n_{R,\vec{k}^{\prime}}^{s^{\prime}\dag}+h.c.)
]\\
=\int{\frac{d^3k}{(2\pi)^3} \frac{d^3k^{\prime}}{(2\pi)^3}}
&[\delta(\vec{k}-\vec{k}^{\prime}) \cdot k_i 
\sum_{s,s^{\prime}}(
e_{\vec{k}}^{s\dag}e_{\vec{k}^{\prime}}^{s^{\prime}}
n_{L,\vec{k}}^{s\dag} \bar{\sigma}^i n_{L,\vec{k}^{\prime}}^{s^{\prime}}
-e_{\vec{k}}^{s}e_{\vec{k}^{\prime}}^{s^{\prime}\dag}
n_{R,\vec{k}}^s \sigma^i n_{R,\vec{k}^{\prime}}^{s^{\prime}\dag})\\
&+\tilde{m}_N \delta(-\vec{k}-\vec{k}^{\prime})
\sum_{s,s^{\prime}}(
e_{\vec{k}}^{s\dag} e_{\vec{k}^{\prime}}^{s^{\prime}\dag}
n_{L,\vec{k}}^{s\dag} n_{R,\vec{k}^{\prime}}^{s^{\prime}\dag} + h.c.
)]\\
=\int{\frac{d^3k}{(2\pi)^3} }
\sum_s
&[k_i (n_{L,\vec{k}}^{s\dag} \bar{\sigma}^i n_{L,\vec{k}}^s
- n_{R,\vec{k}}^s \sigma^i n_{R,\vec{k}}^{s\dag})
+\tilde{m}_N ( se^{i\theta_{\vec{k}}} n_{L,\vec{k}}^{s\dag} n_{R,-\vec{k}}^{s\dag}
+se^{-i\theta_{\vec{k}}} n_{R,-\vec{k}}^{s} n_{L,\vec{k}}^{s}
)]\\
=\int{\frac{d^3k}{(2\pi)^3}} \sum_s
&[-s \lvert\vec{k}\rvert (n_{L,\vec{k}}^{s\dag} n_{L,\vec{k}}^s
+ n_{R,\vec{k}}^s n_{R,\vec{k}}^{s\dag})
+\tilde{m}_N ( se^{i\theta_{\vec{k}}} n_{L,\vec{k}}^{s\dag} n_{R,-\vec{k}}^{s\dag}
+se^{-i\theta_{\vec{k}}} n_{R,-\vec{k}}^{s} n_{L,\vec{k}}^{s}
)]\,.  
\label{Hamiltonian}
\end{aligned}
\end{equation}
Here we have used the orthogonality relations:
\begin{align}
e_{\vec{k}}^{s\dag} e_{\vec{k}}^{s^{\prime}}&=\delta^{s s^{\prime}}\,, \notag \\
e_{\vec{k}}^{s} e_{\vec{k}}^{s^{\prime}\dag}&=\delta^{s s^{\prime}}\,, \notag\\ 
e_{\vec{k}}^{s\dag} e_{-\vec{k}}^{s^{\prime}\dag} 
&=\delta_{s s^{\prime}} \cdot s e^{i\theta_{\vec{k}}}\,, \notag \\
e_{-\vec{k}}^s e_{\vec{k}}^{s^{\prime}}
&=\delta_{s s^{\prime}} \cdot s e^{-i\theta_{\vec{k}}} 
\,.
\end{align} 
Thus the kinetic energy density operator, normalized by the volume factor $V$,  
is read off as 
\begin{equation}
\rho_{\vec{k}}=\frac{1}{V}\sum_{s=\pm}
[-s \lvert\vec{k}\rvert
(n_{L,\vec{k}}^{s\dag} n_{L,\vec{k}}^s
+ n_{R,\vec{k}}^s n_{R,\vec{k}}^{s\dag})
+\tilde{m}_N ( se^{i\theta_{\vec{k}}} n_{L,\vec{k}}^{s\dag} n_{R,-\vec{k}}^{s\dag}
+se^{-i\theta_{\vec{k}}} n_{R,-\vec{k}}^{s} n_{L,\vec{k}}^{s}
)]\,,  
\end{equation}
which reproduces Eq.(\ref{energy_density}).

\section{Relevant time-evolution equations for two-point functions among spinor fields and setting of those initial conditions} 
\label{two-point}

From Eq.(\ref{EOMs}) the time-evolution equations relevant to 
the BNA $n_B(t)$ in Eq.(\ref{nB}) are derived as follows: 
\begin{align}
{\cal D}_t \langle \nLkd \nLk \rangle  
&= 
-is \eiqk 
\left( 
\tilmN \langle \nRmk \nLk \rangle  
+ 
g 
\langle \DRmk \nLk \rangle  
\right)
+ {\rm h.c.}\,,\\
{\cal D}_t \langle \nRmk \nLk \rangle  
&= 
2 i s \abk \langle \nRmk \nLk \rangle  
- 
is \eiqmk \left( \tilmN \langle \nLkd \nLk \rangle  + g \langle \DLkd \nLk \rangle  \right) \notag\\
&\hspace{5mm}+ 
is \eiqmk \left( \tilmN \langle \nRmk \nRmkd \rangle  + g \langle \nRmk \DRmkd \rangle  \right)\,,\\
{\cal D}_t \langle \DRmk \nLk \rangle  
&= 
2is \abk \langle \DRmk \nLk \rangle  
+ 
is \eiqmk \left( \tilmN \langle \DRmk \nRmkd \rangle  + g\langle \DRmk \DRmkd \rangle  \right) \notag\\
&\hspace{5mm}- 
is \eiqmk \left( m_\chi \langle \DLkd \nLk \rangle  + g \langle \nLkd \nLk \rangle  \right)\,,\\
{\cal D}_t \langle \DLkd \nLk \rangle  
&= 
is \eiqmk \left( \tilmN \langle \DLkd \nRmkd \rangle  + g \langle \DLkd \DRmkd \rangle  \right) \notag\\
&\hspace{5mm}- 
is \eiqk \left( m_\chi \langle \DRmk \nLk \rangle   + g \langle \nRmk \nLk \rangle  \right)\,,\\
{\cal D}_t \langle \nRmkd \nRmk \rangle 
&= 
is \eiqk \left( \tilmN \langle \nLk \nRmk \rangle  + g \langle \DLk \nRmk \rangle  \right) 
+ {\rm h.c.}\,,\\
{\cal D}_t \langle \DRmkd \nRmk \rangle  
&= 
is \eiqk \left( m_\chi \langle \DLk \nRmk \rangle    + g \langle \nLk \nRmk \rangle  \right) \notag\\
&\hspace{5mm}- 
is \eiqmk \left( \tilmN \langle \DRmkd \nLkd \rangle  + g \langle \DRmkd \DLkd \rangle  \right)\,,\\
{\cal D}_t \langle \DRmkd \DRmk \rangle  
&= 
is \eiqk \left( m_\chi \langle \DLk \DRmk \rangle   + g \langle \nLk \DRmk \rangle  \right) 
+ {\rm h.c.}\,,\\
{\cal D}_t \langle \DRkd \nLk \rangle  
&= 
-is \eiqk \left( m_\chi \langle \DLmk \nLk \rangle   + g \langle \nLmk \nLk \rangle  \right)\notag\\
&\hspace{5mm}+ 
is \eiqmk \left( \tilmN \langle \DRkd \nRmkd \rangle  + g \langle \DRkd \DRmkd \rangle  \right)\,,\\
{\cal D}_t \langle \DLk \nRmk \rangle  
&= 
2is \abk \langle \DLk \nRmk \rangle  +
is \eiqmk \left( m_\chi \langle \DRmkd \nRmk \rangle     + g \langle \nRmkd \nRmk \rangle  \right)\notag\\
&\hspace{5mm}- 
is \eiqmk \left( \tilmN \langle \DLk \nLkd \rangle  + g \langle \DLk \DLkd \rangle  \right)\,,\\
{\cal D}_t \langle \DRk \nRmk \rangle  
&= 
2is \abk \langle \DRk \nRmk \rangle  -is \eiqmk \left( \tilmN \langle \DRk \nLkd \rangle  + g \langle \DRk \DLkd \rangle  \right) \notag\\
&\hspace{5mm}+ 
is \eiqmk \left( m_\chi \langle \DLmkd \nRmk \rangle     + g \langle \nLmkd \nRmk \rangle  \right)\,, 
\end{align}

\begin{align}
{\cal D}_t \langle \DRmk \DLk \rangle  
&= 
2is \abk \langle \DRmk \DLk \rangle  -is \eiqmk \left( m_\chi \langle \DLkd \DLk \rangle     + g \langle \nLkd \DLk \rangle  \right)\notag\\
&\hspace{5mm}+ 
is \eiqmk \left( m_\chi \langle \DRmk \DRmkd \rangle   + g \langle \DRmk \nRmkd \rangle  \right)\,,\\
{\cal D}_t \langle \DLmk \nLk \rangle  
&= 
2is \abk \langle \DLmk \nLk \rangle  + is \eiqmk \left( \tilmN \langle \DLmk \nRmkd \rangle  + g \langle \DLmk \DRmkd \rangle  \right)\notag\\
&\hspace{5mm}- 
is \eiqmk \left( m_\chi \langle \DRkd \nLk \rangle  + g \langle \nRkd \nLk \rangle  \right)\,,\\
{\cal D}_t \langle \nLmk \nLk \rangle  
&= 
2is \abk \langle \nLmk \nLk \rangle  - is \eiqmk \left( \tilmN \langle \nRkd \nLk \rangle  + g \langle \DRkd \nLk \rangle  \right)\notag\\
&\hspace{5mm}+ 
is \eiqmk \left( \tilmN \langle \nLmk \nRmkd \rangle  + g \langle \nLmk \DRmkd \rangle  \right)\,,\\
{\cal D}_t \langle \DRk \DRmk \rangle  
&= 
2is \abk \langle \DRk \DRmk \rangle  - is \eiqmk \left( m_\chi \langle \DRk \DLkd \rangle    + g \langle \DRk \nLkd \rangle  \right)\notag\\
&\hspace{5mm} +
is \eiqmk \left( m_\chi \langle \DLmkd \DRmk \rangle   + g \langle \nLmkd \DRmk \rangle  \right)\,,\\
{\cal D}_t \langle \DLmkd \nRmk \rangle  
&= 
is \eiqk \left( m_\chi \langle \DRk \nRmk \rangle    + g \langle \nRk \nRmk \rangle  \right)\notag\\
&\hspace{5mm}- 
is \eiqmk \left( \tilmN \langle \DLmkd \nLkd \rangle  + g \langle \DLmkd \DLkd \rangle  \right)\,,\\
{\cal D}_t \langle \DLkd \DLk \rangle  
&= 
-is \eiqk \left( m_\chi \langle \DRmk \DLk \rangle  + g \langle \nRmk \DLk \rangle  \right) 
+ {\rm h.c.}\,,\\
{\cal D}_t \langle \nRk \nRmk \rangle  
&= 
2is \abk \langle \nRk \nRmk \rangle  + is \eiqmk \left( \tilmN \langle \nLmkd \nRmk \rangle  + g \langle \DLmkd \nRmk \rangle  \right)\notag\\
&\hspace{5mm}- 
is \eiqmk \left( \tilmN \langle \nRk \nLkd \rangle  + g \langle \nRk \DLkd \rangle  \right)\,,\\
{\cal D}_t \langle \DLmk \DLk \rangle  
&= 
2is \abk \langle \DLmk \DLk \rangle  -is \eiqmk \left( m_\chi \langle \DRkd \DLk \rangle     + g \langle \nRkd \DLk \rangle  \right)\notag\\
&\hspace{5mm}+ 
is \eiqmk \left( m_\chi \langle \DLmk \DRmkd \rangle    + g \langle \DLmk \nRmkd \rangle  \right)\,,\\
{\cal D}_t \langle \DLkd \DRk \rangle  
&= 
is \eiqmk\left( m_\chi \langle \DLkd \DLmkd \rangle  + g \langle \DLkd \nLmkd \rangle  \right)\notag\\
&\hspace{5mm}- 
is \eiqk \left( m_\chi \langle \DRmk \DRk \rangle    + g \langle \nRmk \DRk \rangle  \right)\,,\\
{\cal D}_t \langle \nRkd \nLk \rangle  
&= 
-i s\eiqk \left( \tilmN \langle \nLmk \nLk \rangle  + g \langle \DLmk \nLk \rangle  \right)\notag\\
&\hspace{5mm}+ 
is \eiqmk \left( \tilmN \langle \nRkd \nRmkd \rangle  + g \langle \nRkd \DRmkd \rangle  \right)\,. 
\end{align}

To solve the sets of equations derived as above, we also need to place the initial conditions. 
We expect that the mass-eigenstate fields in the system initially behave as free fields because the fields do not have any other interactions except those due to $\langle \sigma (t) \rangle $.  
In that case, the initial conditions on the mass-eigenstate fields are simply 
fixed through the zero-particle states defined in the conventional static (adiabatic) vacuum.

We introduce a collective two-component spinor field as 
\begin{equation}
\left[\psi_{\vec{k}}^{s }\right]^{i}=   
\left(\begin{array}{c}
n_{L, \vec{k}}^{s}(t) \\
n_{R, \vec{k}}^{s}(t) \\
D_{L, \vec{k}}^{s}(t) \\
D_{R, \vec{k}}^{s}(t)
\end{array}\right)^i
\,.
\end{equation} 
By defining 
\begin{equation}
\begin{aligned}
&{\left[A_{\vec{k}}\right]^{i j} \equiv \frac{1}{2 V} \sum_{s}\left(\left\langle\left[\psi_{\vec{k}}^{s \dagger}\right]^{i}\left[\psi_{\vec{k}}^{s}\right]^{j}\right\rangle -\left\langle\left[\psi_{-\vec{k}}^{s}\right]^{j}\left[\psi_{-\vec{k}}^{s \dagger}\right]^{i}\right\rangle \right)} \\
&{\left[\bar{A}_{\vec{k}}\right]^{i j} \equiv \frac{1}{2 V} \sum_{s} s\left(\left\langle\left[\psi_{\vec{k}}^{s \dagger}\right]^{i}\left[\psi_{\vec{k}}^{s}\right]^{j}\right\rangle -\left\langle\left[\psi_{-\vec{k}}^{s}\right]^{j}\left[\psi_{-\vec{k}}^{s \dagger}\right]^{i}\right\rangle \right)} \\
&{\left[B_{\vec{k}}\right]^{i j} \equiv \frac{1}{2 V} \sum_{s}\left(\left\langle s e^{i \theta_{\vec{k}}}\left[\psi_{-\vec{k}}^{s}\right]^{i}\left[\psi_{\vec{k}}^{s}\right]^{j}\right\rangle +\left\langle s e^{i \theta_{\vec{k}}}\left[\psi_{-\vec{k}}^{s}\right]^{j}\left[\psi_{\vec{k}}^{s}\right]^{i}\right\rangle \right)} \\
&{\left[\bar{B}_{\vec{k}}\right]^{i j} \equiv \frac{1}{2 V} \sum_{s} s\left(\left\langle s e^{i \theta_{\vec{k}}}\left[\psi_{-\vec{k}}^{s}\right]^{i}\left[\psi_{\vec{k}}^{s}\right]^{j}\right\rangle +\left\langle s e^{i \theta_{\vec{k}}}\left[\psi_{-\vec{k}}^{s}\right]^{j}\left[\psi_{\vec{k}}^{s}\right]^{i}\right\rangle \right)}
\end{aligned}
\,,  
\end{equation}  
their initial conditions at $t=t_0$ are then set as~\cite{Enomoto:2022gsx}  
\begin{equation}
\begin{gathered}
{\left[A_{\vec{k}}\left(t_{0}\right)\right]^{i j}=\left[\bar{B}_{\vec{k}}\left(t_{0}\right)\right]^{i j}=0} \,, \\
{\left[\bar{A}_{\vec{k}}\left(t_{0}\right)\right]^{i j}=\sum_{\ell}\left[U^{*}\right]^{i \ell} \frac{|\vec{k}|}{\omega_{k}^{\ell}}\left[U^{T}\right]^{\ell j}} \,, \\
{\left[B_{\vec{k}}\left(t_{0}\right)\right]^{i j}=\sum_{\ell} [U]^{i \ell} \frac{\mathcal{M}_{d}^{\ell \ell}}{\omega_{k}^{\ell}}\left[U^{T}\right]^{\ell j}}
\,, 
\end{gathered}
\label{initial:condi}
\end{equation}
    where $\mathcal{M}_{d}$ is a diagonalized mass matrix related to the mass matrix $\mathcal{M}$ 
    through a (orthogonal) transformation by $U$, which we have performed the Takagi factorization method: 
\begin{align}
\mathcal{M} &= \left( 
\begin{array}{cccc} 
0 & \tilde{m}_N & 0 & g \\ 
\tilde{m}_N & 0 & g & 0 \\ 
0 & g & 0 & m_\chi \\ 
g & 0 & m_\chi & 0  
\end{array} 
\right)\,, \notag\\ 
\mathcal{M}_{d} 
&\equiv U^{T} \mathcal{M} U 
\,. 
\end{align}
In Eq.(\ref{initial:condi}) $\omega_{k}^{\ell}$ is the energy with the diagonalized mass as
\begin{equation}
    \omega_{k}^{\ell} \equiv \sqrt{|\vec{k}|^{2}+\left(\mathcal{M}_{d}^{\ell \ell}\right)^{2}} .
\end{equation}

In terms of the two-point functions, the initial conditions (set at $t_0=0$) read~\cite{Enomoto:2022gsx} 
\begin{equation}
\begin{array}{lll}
\left\langle O_{1}\right\rangle_{t=0} & =\left\langle n_{L, \vec{k}}^{s \dagger} n_{L, \vec{k}}^{s}\right\rangle_{t=0} & =\frac{V}{2}\left(1+s\left[\bar{A}_{k}\left(t_{0}\right)\right]^{11}\right)\,,  \\
\left\langle O_{2}\right\rangle_{t=0} & =\left\langle n_{R,-\vec{k}}^{s} n_{L, \vec{k}}^{s}\right\rangle_{t=0} & =\frac{V s}{2}\left[B_{k}\left(t_{0}\right)\right]^{12}\,,  \\
\left\langle O_{3}\right\rangle_{t=0} & =\left\langle D_{R,-\vec{k}}^{s} n_{L, \vec{k}}^{s}\right\rangle_{t=0} & =\frac{V s}{2}\left[B_{k}\left(t_{0}\right)\right]^{41}\,,  \\
\left\langle O_{4}\right\rangle_{t=0} & =\left\langle D_{L, \vec{k}}^{s \dagger} n_{L, \vec{k}}^{s}\right\rangle_{t=0} & =\frac{V}{2}\left(s\left[\bar{A}_{k}\left(t_{0}\right)\right]^{31}\right)\,,  \\
\left\langle O_{5}\right\rangle_{t=0} & =\left\langle n_{R,-\vec{k}}^{s \dagger} n_{R,-\vec{k}}^{s}\right\rangle_{t=0} & =\frac{V}{2}\left(1+s\left[\bar{A}_{k}\left(t_{0}\right)\right]^{22}\right)\,,  \\
\left\langle O_{6}\right\rangle_{t=0} & =\left\langle D_{R,-\vec{k}}^{s \dagger} n_{R,-\vec{k}}^{s}\right\rangle_{t=0} & =\frac{V}{2}\left(s\left[\bar{A}_{\boldsymbol{k}}\left(t_{0}\right)\right]^{42}\right)\,,  \\
\left\langle O_{7}\right\rangle_{t=0} & =\left\langle D_{R,-\vec{k}}^{s+} D_{R,-\vec{k}}^{s}\right\rangle_{t=0} & =\frac{V}{2}\left(1+s\left[\bar{A}_{k}\left(t_{0}\right)\right]^{44}\right)\,,  \\
\left\langle O_{8}\right\rangle_{t=0} & =\left\langle D_{R, \vec{k}}^{s t} n_{L, \vec{k}}^{s}\right\rangle_{t=0} & =\frac{V}{2}\left(s\left[\bar{A}_{k}\left(t_{0}\right)\right]^{41}\right)\,,  \\
\left\langle O_{9}\right\rangle_{t=0} & =\left\langle D_{L, \vec{k}}^{s} n_{R,-\vec{k}}^{s}\right\rangle_{t=0} & =-\frac{V s}{2}\left[B_{k}\left(t_{0}\right)\right]^{32}\,,  \\ 
\left\langle O_{10}\right\rangle_{t=0} & =\left\langle D_{R, \vec{k}}^{s} n_{R,-\vec{k}}^{s}\right\rangle_{t=0} & =-\frac{V s}{2}\left[B_{k}\left(t_{0}\right)\right]^{42}\,,  \\
\left\langle O_{11}\right\rangle_{t=0} & =\left\langle D_{R,-\vec{k}}^{s} D_{L, \vec{k}}^{s}\right\rangle_{t=0} & =\frac{V s}{2}\left[B_{k}\left(t_{0}\right)\right]^{34}\,,  \\
\left\langle O_{12}\right\rangle_{t=0} & =\left\langle D_{L,-\vec{k}}^{s} n_{L, \vec{k}}^{s}\right\rangle_{t=0} & =\frac{V s}{2}\left[B_{k}\left(t_{0}\right)\right]^{31}\,,  \\
\left\langle O_{13}\right\rangle_{t=0} & =\left\langle n_{L,-\vec{k}}^{s} n_{L, \vec{k}}^{s}\right\rangle_{t=0} & =\frac{V s}{2}\left[B_{k}\left(t_{0}\right)\right]^{11}\,,  \\
\left\langle O_{14}\right\rangle_{t=0} & =\left\langle D_{R, \vec{k}}^{s} D_{R,-\vec{k}}^{s}\right\rangle_{t=0} & =-\frac{V s}{2}\left[B_{k}\left(t_{0}\right)\right]^{44}\,,  \\
\left\langle O_{15}\right\rangle_{t=0} & =\left\langle D_{L,-\vec{k}}^{s \dagger} n_{R,-\vec{k}}^{s}\right\rangle_{t=0} & =\frac{V}{2}\left(s\left[\bar{A}_{k}\left(t_{0}\right)\right]^{32}\right)\,,  \\
\left\langle O_{16}\right\rangle_{t=0} & =\left\langle D_{L, \vec{k}}^{s \dagger} D_{L, \vec{k}}^{s}\right\rangle_{t=0} & =\frac{V}{2}\left(1+s\left[\bar{A}_{k}\left(t_{0}\right)\right]^{33}\right)\,,  \\
\left\langle O_{17}\right\rangle_{t=0} & =\left\langle n_{R, \vec{k}}^{s} n_{R,-\vec{k}}^{s}\right\rangle_{t=0} & =-\frac{V s}{2}\left[B_{k}\left(t_{0}\right)\right]^{22}\,,  \\
\left\langle O_{18}\right\rangle_{t=0} & =\left\langle D_{L,-\vec{k}}^{s} D_{L, \vec{k}}^{s}\right\rangle_{t=0} & =\frac{V s}{2}\left[B_{k}\left(t_{0}\right)\right]^{33}\,,  \\
\left\langle O_{19}\right\rangle_{t=0} & =\left\langle D_{L, \vec{k}}^{s \dagger} D_{R, \vec{k}}^{s}\right\rangle_{t=0} & =\frac{V}{2}\left(s\left[\bar{A}_{k}\left(t_{0}\right)\right]^{34}\right)\,,  \\
\left\langle O_{20}\right\rangle_{t=0} & =\left\langle n_{R, \vec{k}}^{s \dagger} n_{L, \vec{k}}^{s}\right\rangle_{t=0} & =\frac{V}{2}\left(s\left[\bar{A}_{k}\left(t_{0}\right)\right]^{12}\right)^{*}\,.
\end{array}
\end{equation}

\end{document}